\documentclass[aps,preprint,pra]{revtex4-1}

\usepackage{graphicx}
\usepackage{amsfonts}
\usepackage{amsmath}
\usepackage{caption}
\usepackage{subcaption}
\usepackage[super]{nth}

\newcommand{\bra}[1]{{\langle}#1|}
\newcommand{\ket}[1]{|#1{\rangle}}

\newcommand{\sinc}{\operatorname{sinc}}

\begin{document}

\title{Self-ordering dynamics of ultracold atoms in multicolored cavity fields}

\author{S. Kr\"amer}
\affiliation{Institute for Theoretical Physics, Universit\"at Innsbruck, Technikerstrasse 25, 6020 Innsbruck, Austria}
\author{H. Ritsch}
\affiliation{Institute for Theoretical Physics, Universit\"at Innsbruck, Technikerstrasse 25, 6020 Innsbruck, Austria}

\begin{abstract}
We study light induced spatial crystallization of ultracold quantum particles confined along the axis of a high-$Q$ linear cavity via a transverse multicolor pump using numerical simulations. Whenever a pump frequency is tuned close to resonance with a longitudinal cavity mode, the dynamics favors bistable spatial particle ordering into a Bragg grating at a wavelength distance. Simultaneous pumping at several resonant frequencies fosters competition between the different spatial lattice orders, exhibiting complex nonlinear field dynamics involving several metastable atom-field states. For few particles even superpositions of different spatial orders entangled with different light mode amplitudes appear. By a proper choice of trap geometry and pump frequencies a broad variety of many particle Hamiltonians with a nontrivial long range coupling can be emulated in such a setup. When applying quantum Monte Carlo wave function simulations to study time evolution we find simultaneous super radiant scattering into several light modes and the buildup of strong non-classical atom field correlations.
\end{abstract}

\maketitle

\section{Introduction}
When ultracold polarizable particles in an optical resonator are coherently illuminated from the side at sufficient intensity, the system undergoes a transition from homogeneous to crystalline order accompanied by superradiant light scattering into the cavity mode~\cite{domokos2002collective,black2003observation,arnold2012self}. This self ordering process occurs as a quantum phase transition for an ultracold Bose as well as Fermi gas close to zero temperature in the pure quantum regime~\cite{maschler2007entanglement,piazza2013umklapp}. To surprising accuracy it can be reinterpreted and understood as implementation of the Dicke superradiant phase transition, where a cavity mode is very strongly coupled to two motional modes of the trapped particles~\cite{baumann2010dicke,nagy2010dicke, ritsch2013cold}. As one of the early and seminal examples of a quantum phase transition~\cite{hepp1973superradiant,sachdev2007quantum}, the Dicke superradiant phase transition has been studied extensively in theory since its proposition 50 years ago~\cite{hepp1973superradiant}. With the recent successful experimental implementation based on a BEC of cold atoms in a cavity, a large number of more detailed theoretical and numerical studies on the particular properties of this realization, its limitations and new related phenomena like super-solidity have followed~\cite{strack2011dicke,tolkunov2006quantum,piazza2013bose,konya2012finite,li2013lattice}.

The first simple but still surprisingly successful generation of theoretical models was based on a single atomic and a single cavity mode~\cite{nagy2010dicke}. A particular interesting generalization was introduced by considering this ordering phenomenon in degenerate multi-mode cavity fields and restricting the degrees of freedom of the particle motion to a 2D plane~\cite{gopalakrishnan2009emergent}. This opened a new possible route to study a wealth of important quantum solid state phenomena like emergent crystallinity, liquid crystalline phases~\cite{gopalakrishnan2009emergent,gopalakrishnan2010viewpoint} or quantum neural network models~\cite{gopalakrishnan2012exploring} in a well controllable atom-optical configuration. Due to the inherent complexity and technical difficulty of degenerate multi-mode setups, which require at least a 2D geometry, theoretical modeling as well as experimental implementations of such setups are very challenging and have, to our knowledge, not been implemented to date. Recent theoretical generalizations to fermions reveal even more new physical phenomena, e.g.\ Umklapp lasing or surprisingly reduced ordering thresholds~\cite{piazza2013umklapp,keeling2013fermionic,chen2013superradiance}.

In this work we investigate an alternative multi-mode extension employing several light frequencies simultaneously, each of them tuned closely to a separate longitudinal cavity mode. As the cavity modes form an equidistant comb of sufficiently distinct frequencies, which is readily available experimentally with modern comb technology, the technical complexity of an implementation seems much less challenging. Moreover, as coherent light scattering between different modes can be neglected and the motion of the particles is confined along the cavity axis, the computational complexity of the model is substantially reduced. This allows explicit numerical treatments, at least for small implementations, where still many of the intriguing phenomena mentioned above should appear. As amplitude, phase and detuning of each mode can be controlled independently, this setup should allow for much more precise control and variability of the dynamics.

While virtually all the existing treatments have ignored the effect of the finite size of the particle trap inside the resonator, here, we explicitly consider particles confined to a finite volume along the cavity axis. Hence, the ground state is not translation invariant and neither are the higher trap eigenmodes which form the ordered pattern. This implies a breakdown of the single mode model for a monochromatic pump. In the multicolor pump case, this induces important extra couplings between many different trap modes, which can be tailored largely by the choice of the pump and trap geometry. In any case, the relevant particle Hilbert space is substantially enlarged.

We will first present the theoretical model assuming the particles to be trapped in a 1D external trap along the cavity axis. For a few generic trap configurations, the scattering and coupling integrals between the light field modes and trapped particle modes can even be calculated analytically, a numerical evaluation for more realistic trap geometries is straightforward. Ultimately, we obtain a well defined effective Hamiltonian consisting of many modes with widely tunable coupling elements. Within this framework at first, we shortly review the single frequency phase transition but account for the full trap dynamics beyond a single excited trap mode. In the last section we increase the number of pump fields and simulate the phase transition and steady states for a multicolor setup.

\section{Model}
\begin{figure}[ht]
    \includegraphics[width=0.4\textwidth]{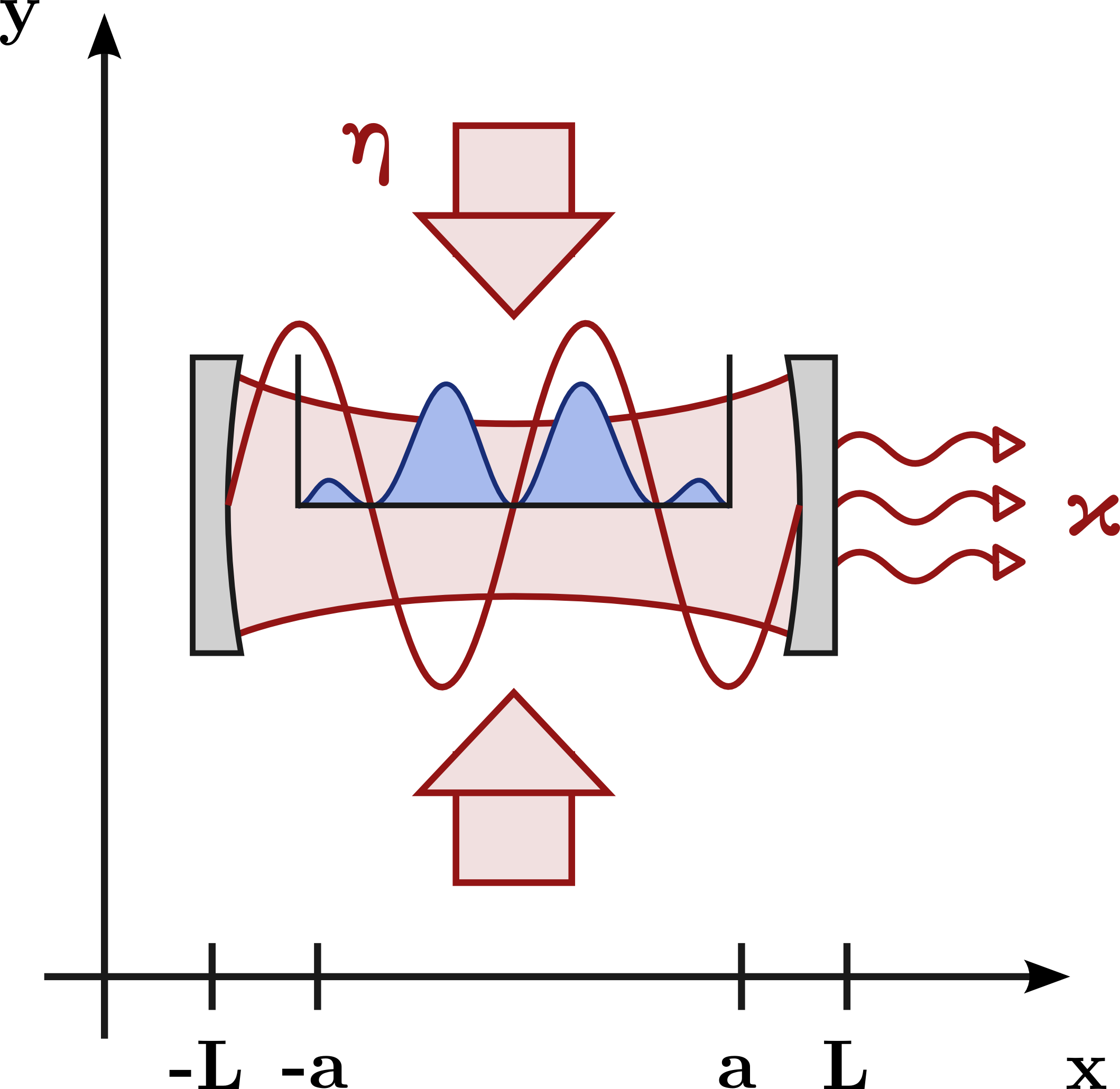}
  \caption{System setup consisting of $N_a$ quantum particles confined in a box trap $V(x)$ along the cavity axis and illuminated by $N_p$ lasers along the y-direction.}
\label{img:cavity-box-potential}
\end{figure}

We consider a linear cavity of length $2 L$ with $N_a$ particles trapped by an external potential $V(x)$ in a finite region along the cavity axis as depicted in fig.~\ref{img:cavity-box-potential}. For simplicity we assume the particles to be tightly confined in the radial direction, making their movement effectively one dimensional.
The particles are characterized solely by their center of mass motion and their electric dipole moment, which might vary strongly with frequency, ultimately depending on which type of particles are used. They are pumped transversely (in y-direction) by $N_p$ different lasers, creating standing waves polarized orthogonally to the cavity with pump strength $E_p^n$ and frequency $\omega_p^n$, where each of the frequencies is tuned closely to a distinct cavity eigenmode of frequency $\omega_c^n$.
Taking spontaneous emission of the cavity modes into account the time evolution of the coupled particle-field system can be described by a master equation of the form~\cite{asboth2005self,ritsch2013cold}
\begin{equation}
  \dot{\rho} = -\frac{i}{\hbar} [H,\rho] + \sum_n \kappa_n (2 \hat{a}_n \rho \hat{a}_n^\dagger - \rho \hat{a}_n^\dagger \hat{a}_n - \hat{a}_n^\dagger \hat{a}_n \rho),
\label{eq:model-masterequation}
\end{equation}
where the Hamiltonian is given by
\begin{equation}
  H = \sum_{n} \hbar \omega_c^n \hat{a}_n^\dagger \hat{a}_n + \sum_{i=1}^{N_{part}} (\frac{\hat{p}_i^2}{2 m} + V(\hat{x}_i)) - \sum_{i=1}^{N_{part}} \hat{\vec{d}}_i \hat{\vec{E}}(\hat{x_i}).
  \label{eq:model-hamiltonian}
\end{equation}
The electric field in the interaction term, $\hat{\vec{E}}(\hat{x}_i) = \hat{\vec{E}}_c(\hat{x}_i) + \vec{E}_p$, consists of the pump lasers $\vec{E}_p$ modeled as classical standing wave fields of the form $\vec{E}_p = \vec{e}_z \sum_n E_0^n \cos(\omega_p^n t)$ and the cavity field $\hat{\vec{E}}_c(\hat{x}_i) = \vec{e}_z \sum_{n} \sqrt{\frac{\hbar \omega_c^n}{2L\epsilon_0}} u_n(\hat{x}_i)a_n + \mathrm{h.c.}$.
For a planar cavity the effective mode functions on the axis are the trigonometric functions $u_n = \sin(k_n (x+L))$, with wave numbers $k_n$.
After changing into an interaction picture with respect to the operator $\sum_n \omega_p^n a_n^\dagger a_n$ and elimination of fast rotating terms the Hamiltonian reduces to
\begin{align}
  H = &-\sum_n \Delta_c^n \hat{a}_n^\dagger \hat{a}_n + \sum_i (\frac{p_i^2}{2\mu} + V(\hat{x}_i))
  \nonumber
  \\
  &+ \sum_{ni} U_0^n \sin^2(k_n (\hat{x}_i-L)) \hat{a}_n^\dagger \hat{a}_n + \sum_{ni} \eta_n \sin(k_n (\hat{x}_i-L)) (\hat{a}_n^\dagger + \hat{a}_n)
\end{align}
with $\Delta_c^n = \omega_p^n - \omega_c^n$. The exact expressions for the effective pump strengths $\eta_n$ as well as $U_0^n$ depend on which type of particles are used. E.g.~in the case of atoms where all pump frequencies are chosen to be not too far from a particular atomic resonance frequency $\omega_a$ these quantities can be calculated analogously to the single pump case~\cite{hechenblaikner1998cooling} to $U_0^n = \frac{\hbar\omega_c^n}{2L\epsilon_0} \frac{\Delta_a^n}{{(\Delta_a^n)}^2+\gamma^2} |d_{eg}|^2$ and $\eta_n = \frac{1}{2} \sqrt{\frac{\hbar\omega_c^n}{2L \epsilon_0}} E_0^n \frac{\Delta_a^n}{{(\Delta_a^n)}^2+\gamma^2} |d_{eg}|^2$ with $\Delta_a^n = \omega_p^n - \omega_a$. More complex setups, where each pump laser frequency is close to a different atomic resonance $\omega_a^n$ could allow for a very fine control of the interaction strength. An alternative choice of using laser frequencies very far from any internal resonances allows to estimate $U_o^n$ from the particles polarizability at this frequency. Corresponding examples are given in ref.~\cite{nimmrichter2010master}. However, from this point on we will ignore the details of the origin of these coupling coefficients and treat them as freely tunable parameters.

For a many particle description it is convenient to use the Hamiltonian in second quantized form,
\begin{align}
  H = &-\sum_n \Delta_c^n \hat{a}_n^\dagger \hat{a}_n + \sum_i E_i \hat{c}_i^\dagger \hat{c}_i + \sum_{nij} U_0^n A_{nij} \hat{c}_i^\dagger \hat{c}_j \hat{a}_n^\dagger \hat{a}_n +\sum_{nij} \eta_n B_{nij} \hat{c}_i^\dagger \hat{c}_j (\hat{a}_n^\dagger + \hat{a}_n),
\label{eq:model-hamiltonian-compact}
\end{align}
where the coupling strength between particle modes and cavity modes is calculated by
\begin{align}
  A^n_{ij} &= \int_{-\infty}^{\infty} \Psi_i^*(x)\Psi_j(x) \sin^2(k_n (x+L)) dx
\label{eq:model-A-coeff}
  \\
  B^n_{ij} &= \int_{-\infty}^{\infty} \Psi_i^*(x)\Psi_j(x) \sin(k_n (x+L)) dx,
\label{eq:model-B-coeff}
\end{align}
with $\Psi_i$ denoting the eigenfunctions of the single particle Hamiltonian $\frac{p^2}{2m}+V(x)$.

The specific physical parameters of any concrete implementation are now contained in the detuning of the pump lasers from their respective cavity mode, the spectrum of trap eigenenergies $E_i$, the effective pump strengths $\eta_n$ and the parameter $U_0^n$, whereas the coupling matrices $A$ and $B$ capture the geometric properties of the system. Variation of the trap and pump geometry results in a very wide range of achievable couplings and thus bears great potential as a very flexible resource for a quantum simulator~\cite{lloyd1996universal}. Note that extra restrictions introduced by replacing the particles by fermions will further expand the available options. The $B$ matrices mediate the driving of the system by collective scattering photons into the cavity modes, whereas the $A$ matrices describe the diffractive scattering of the particles between different motional trap modes by the cavity photons. Via tailoring these matrices the properties of combined particle-field eigenstates can be widely tuned and could provide a unique realization of a quantum spin glass with long-range interaction~\cite{strack2011dicke} or generalized forms of Hopfield networks~\cite{gopalakrishnan2012exploring}. For several generic cases these matrices can be evaluated explicitly and we will discuss typical examples in the next paragraph.

\subsection{Coupling matrices for particles in a square box potential}
We now study basic properties of the above Hamiltonian by means of the generic example of a particle trapped in a box potential of length $2 a$, i.e.~$V(x)=0$ when $x \in [-a,a]$ and $V(x)=\infty$ otherwise. The eigenfunctions are then $\Psi_i(x) = \frac{1}{\sqrt{a}} \sin(K_i (x+a))$ inside the box and $\Psi_i(x) = 0$ outside of it. Here, $K_i = \frac{\pi}{2a}(i+1)$, $E_i = \frac{\hbar^2}{2\mu}{\left(\frac{\pi (i+1)}{2a}\right)}^2 = \frac{\hbar^2}{2\mu} K_i^2$ while the coupling matrices $A^n_{ij}$ and $B^n_{ij}$ are simple integrals of products of harmonic functions and compute to
\begin{align}
  A^n_{ij} &= \frac{1}{2}\delta_{ij} + \frac{{(-1)}^n}{4} \left(f^{\cos}_{i,j,n}+f^{\cos}_{i,j,-n}-f^{\cos}_{i,-j,n}-f^{\cos}_{i,-j,-n}\right)
\label{eq:box-A-coeff}
  \\
  B^n_{ij} &= \frac{1}{2}\left(-f^{\sin}_{i,j,n}+f^{\sin}_{-i,j,n}+f^{\sin}_{i,-j,n}+f^{\sin}_{i,j,-n}\right)
\label{eq:box-B-coeff}
\end{align}
with $f^{\cos}_{i,j,n} = \sinc \left(\frac{\pi}{2}(i+j+2\frac{a}{L}n)\right) \cos \left(\frac{\pi}{2}(i+j)\right)$ and $f^{\sin}_{i,j,n} = \sinc \left(\frac{\pi}{2}(i+j+\frac{a}{L}n)\right) \sin \left(\frac{\pi}{2}(i+j+n)\right)$.
For a careful choice of trap length $a$ and selection of cavity modes, e.g.\ $a=L/4$ and $n \in \{11,19,27\}$, there are only few nonzero elements, while in general virtually all trap modes can be coupled. For our numerical simulations we will mainly use the two cavity eigenmodes $n=11$ and $n=19$ with $a=L/4$ as typical examples.
\begin{figure}[ht]
  \begin{subfigure}[b]{0.45\textwidth}
    \includegraphics[width=1.0\textwidth]{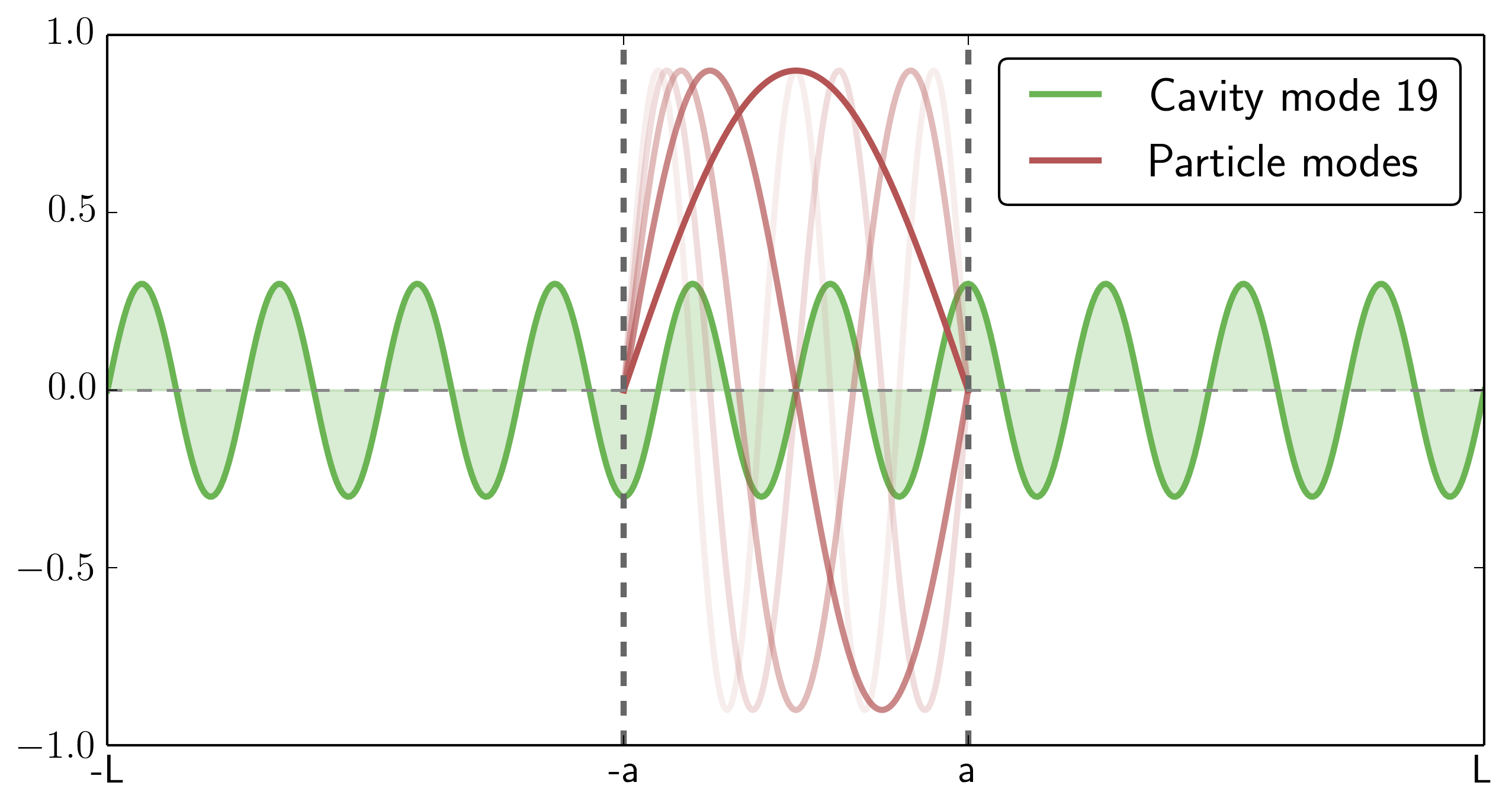}
    (a)
  \end{subfigure}
  \begin{subfigure}[b]{0.5\textwidth}
    \includegraphics[width=1.0\textwidth]{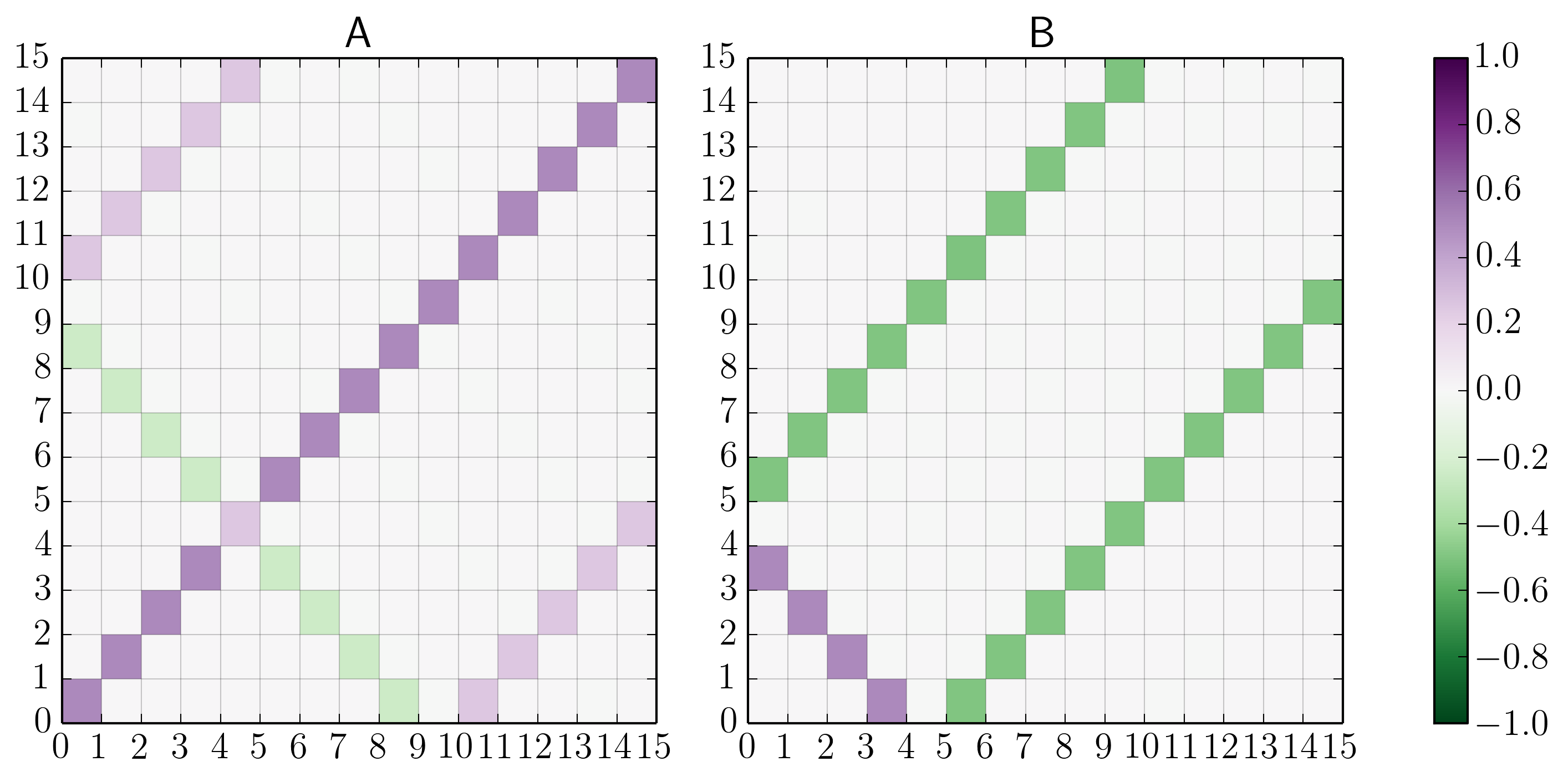}
    (b)
  \end{subfigure}
  \caption{(a) First few particle trap eigenmodes and \nth{19} cavity mode. (b) Coupling matrices $\textrm{A}$ and $\textrm{B}$ between light field and particle modes for the \nth{19} cavity mode.}
\label{img:mode19-conf}
\end{figure}
A closer look at the \nth{19} cavity mode (fig.~\ref{img:mode19-conf}) reveals that only 2 excited particle modes, the \nth{3} and the \nth{5}, are directly coupled to the ground state. The \nth{8} and \nth{10} state are coupled indirectly via these two states to the ground state whereas many other states have no coupling to the ground state at all, which means they can be excluded in our numerical simulations. Unfortunately, this reduction of the Hilbert space can not be done when additional pump lasers are considered since in that case all particle modes will be coupled in some way to the ground state. Note that in addition to a finite coupling strength energy conservation has to be fulfilled for efficient population transfer. This can be controlled via the detuning between the pumping laser and the cavity mode, $\Delta_c^n$, at least to some extent. Nevertheless, even in this simple generic case a two mode picture per excitation frequency as it appears in an infinite trap size limit cannot be expected.

\subsection{Coupling matrices for a harmonic particle trap}
\begin{figure}[ht]
  \begin{subfigure}[b]{0.45\textwidth}
    \includegraphics[width=1.0\textwidth]{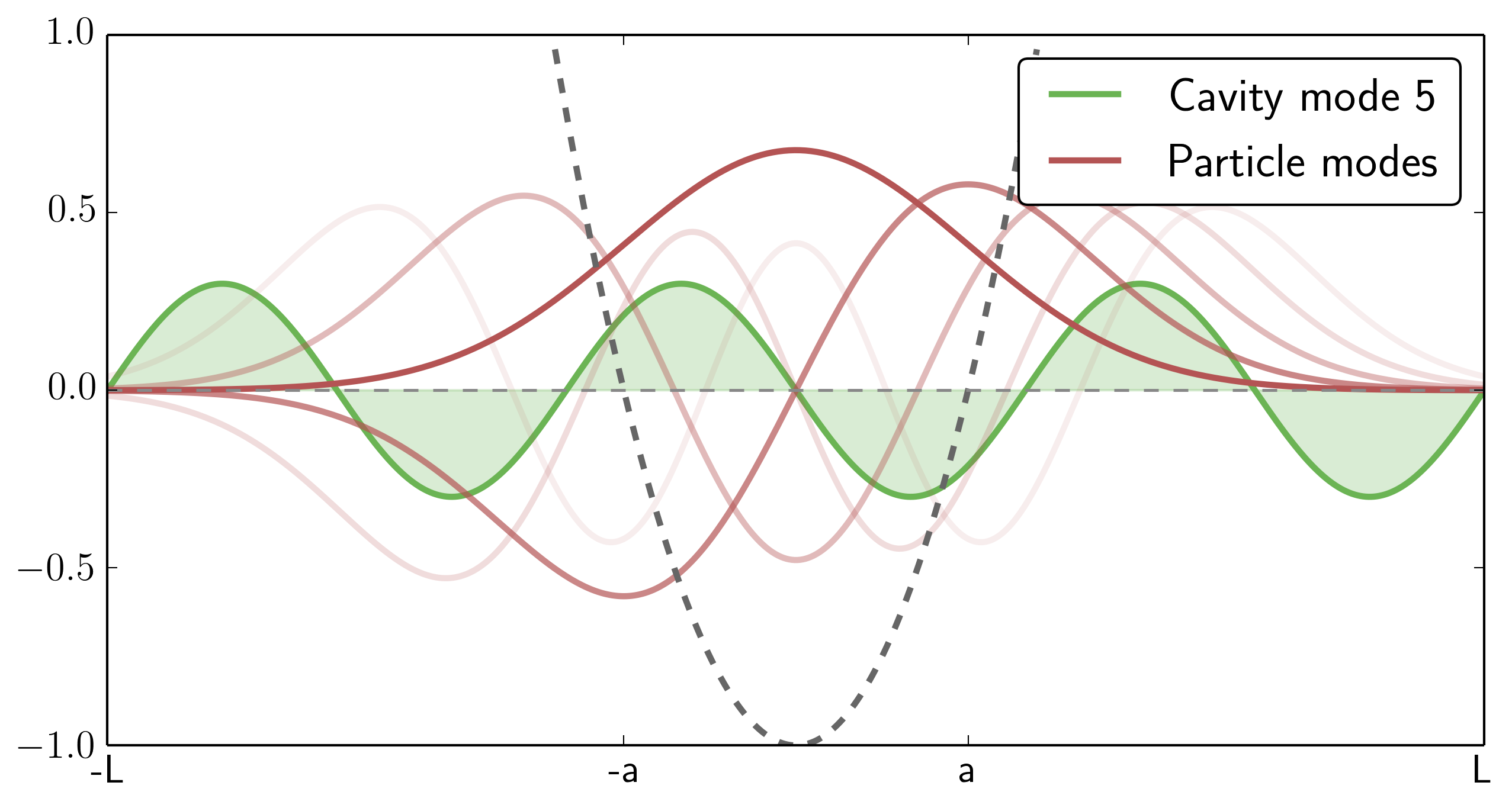}
    (a)
  \end{subfigure}
  \begin{subfigure}[b]{0.5\textwidth}
    \includegraphics[width=1.0\textwidth]{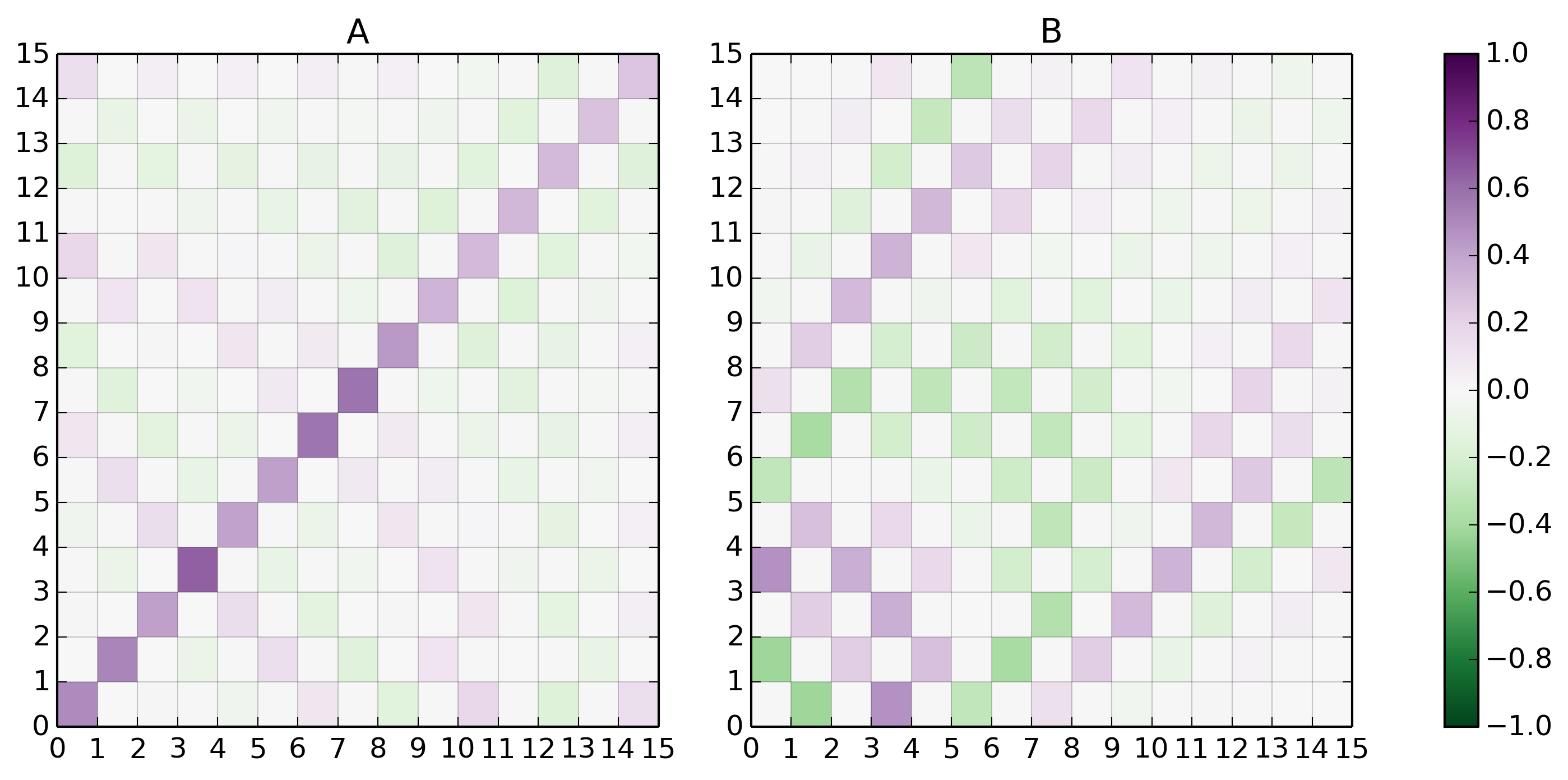}
    (b)
  \end{subfigure}
  \caption{(a) First few particle eigenmodes of an harmonic trap potential and \nth{5} cavity mode. (b) Coupling matrices $\textrm{A}$ and $\textrm{B}$ between light fields and particle modes for a harmonic particle trap positioned at the cavity center.}
\label{img:gaussian-conf}
\end{figure}

For a square-well potential of tailored length as above we find rather sparsely populated coupling matrices with nonzero entries determined by the specific choice of the cavity length, trap size and pump mode. In the more realistic case of e.g.\ a Gaussian trap, the matrices are, as shown in fig.~\ref{img:gaussian-conf}, much more densely populated.
This considerably enlarges the necessary computational basis for quantitatively correct numerical modeling on the one hand, but allows the study of a wider class of Hamiltonians on the other hand. While this is certainly interesting and worthwhile to pursue, we will in the following use the simple, more easily calculable box potential model.

\section{Self-ordering dynamics with a monochromatic pump}
The $B$ matrices calculated above show that photon scattering into the cavity mode couples several motional modes already. For trapped particles, where the translation symmetry of the ground state is lost, there are always at least two motional modes that exhibit a considerable coupling. These motional modes are in turn coupled further to other eigenstates via the $B$ as well as the $A$ matrix. Of course, only if the corresponding motional transition is also energy resonant within the width of the cavity mode, one can expect a significant excitation of a particular mode. Based on this line of argumentation, higher order motional modes have been widely neglected in many theoretical models so far~\cite{konya2011multimode}, even in the bad cavity case.

 In the present work, we treat the more realistic but computationally more challenging case of a finite particle trap placed along the cavity axis, so that we need to include a suitable range of contributing trap eigenstates. As the nonlinear equations cannot be solved analytically, we resort to numerical solutions of the corresponding master equation using the \emph{zvode} wrapper of the numerical library\emph{scipy}~\cite{scipy}. For higher dimensional problems, as it is mostly the case if more than one cavity mode is pumped, we have to settle for QMCWF methods which makes it harder to obtain accurate values for entanglement and correlations but in return allows us to study the dynamics of single trajectories which experimentally corresponds to continuous measurement of leaking photons. These trajectories are often easier to comprehend intuitively, mostly due to the fact that they work in terms of wave functions instead of density matrices, and therefore allow for a better qualitative insight into the underlying processes.
\begin{figure}[ht]
  \begin{subfigure}[b]{0.24\textwidth}
    \includegraphics[width=1.0\textwidth]{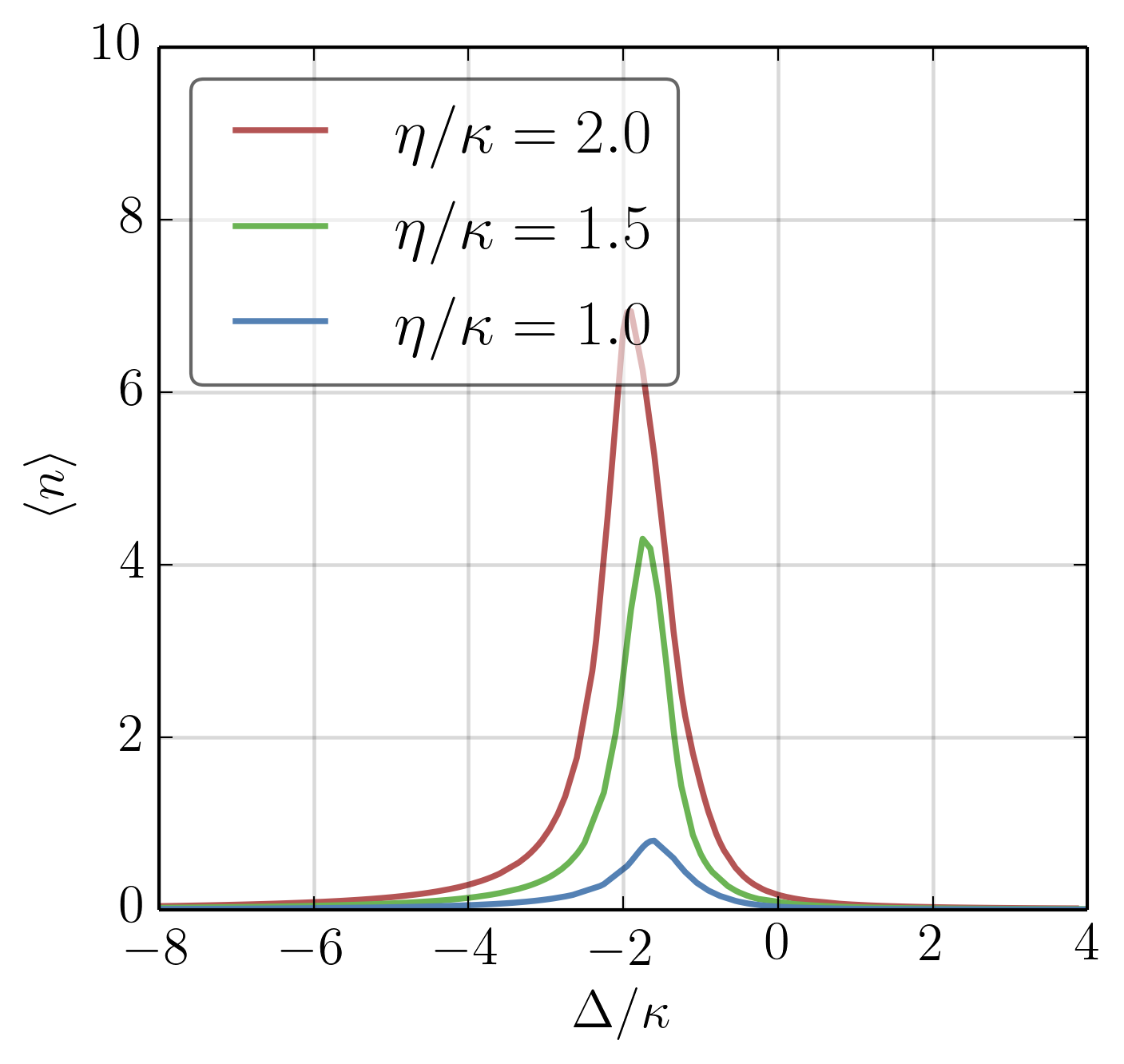}
    (a)
  \end{subfigure}
  \begin{subfigure}[b]{0.24\textwidth}
    \includegraphics[width=1.0\textwidth]{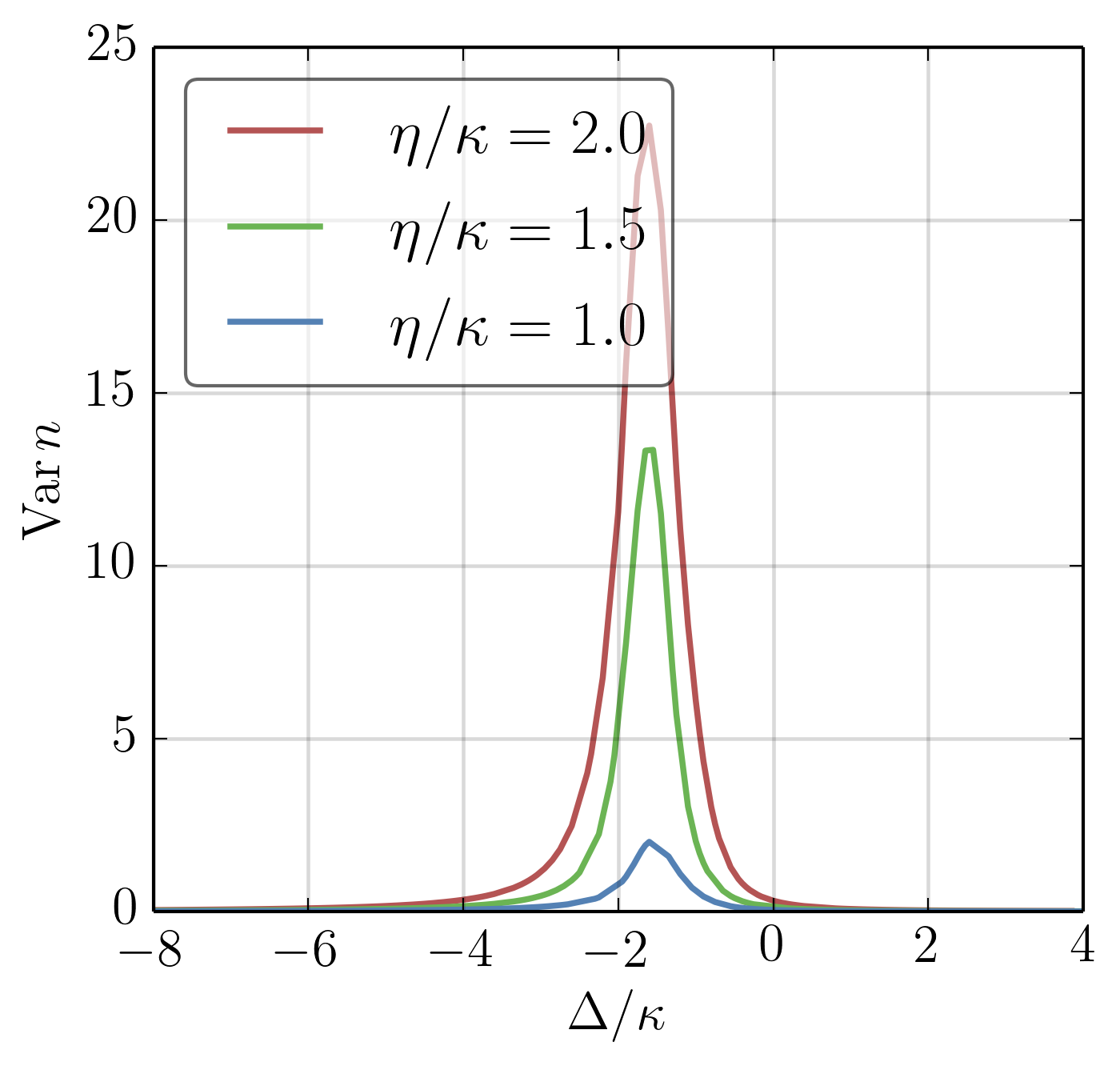}
    (b)
  \end{subfigure}
  \begin{subfigure}[b]{0.24\textwidth}
    \includegraphics[width=1.0\textwidth]{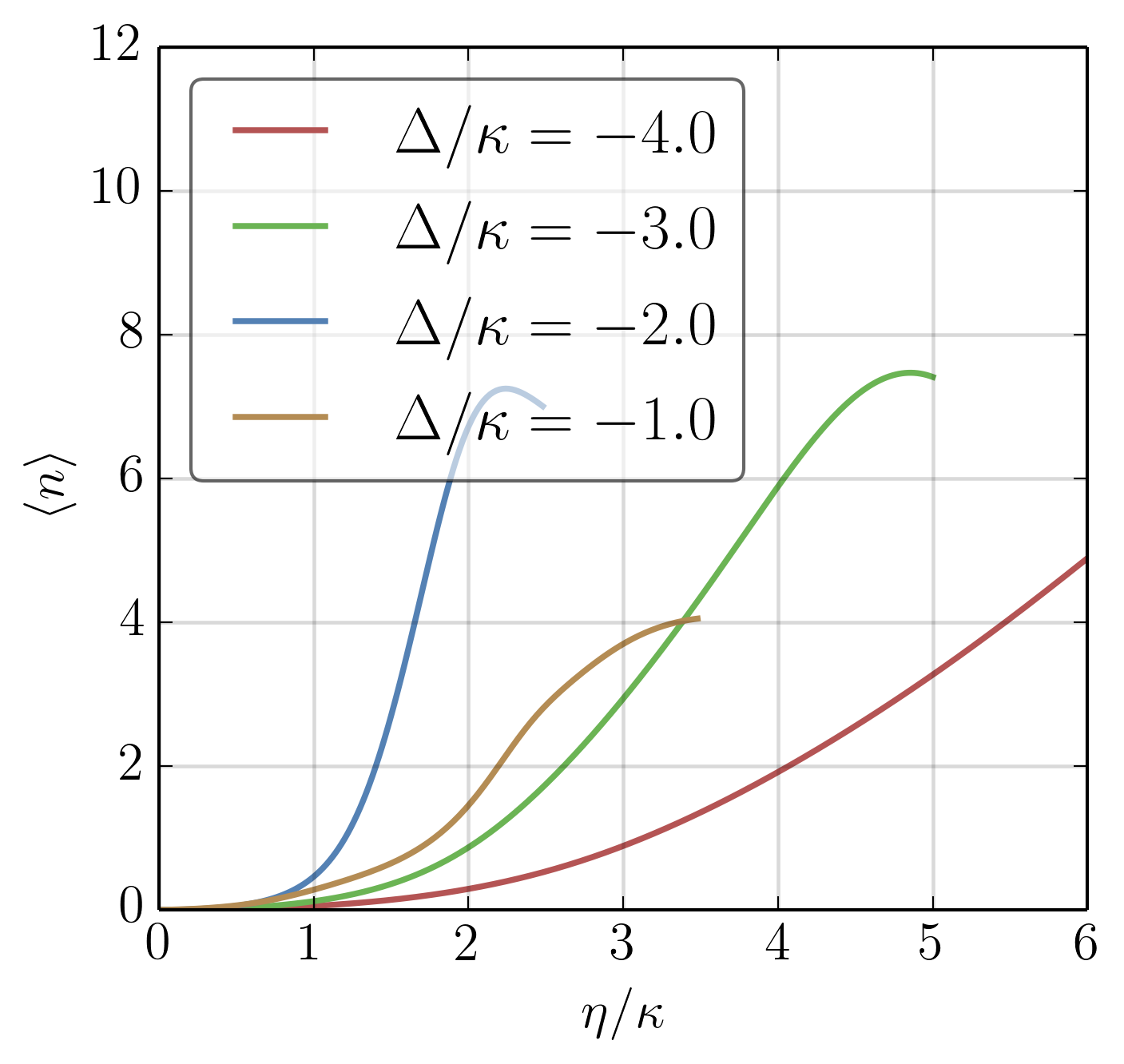}
    (c)
  \end{subfigure}
  \begin{subfigure}[b]{0.24\textwidth}
    \includegraphics[width=1.0\textwidth]{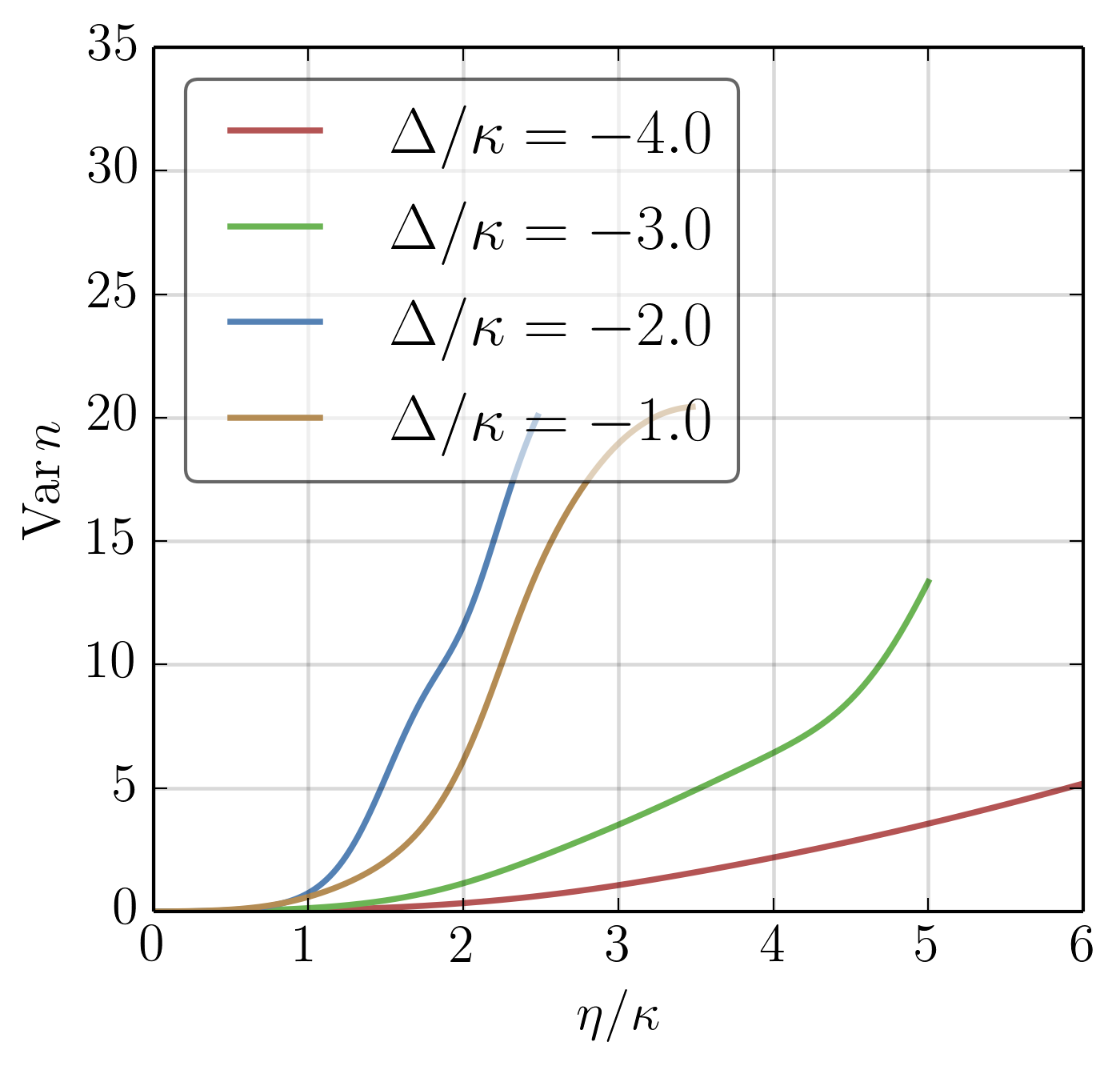}
    (d)
  \end{subfigure}
  \caption{Stationary average number (a) and variance (b) of photons scattered into the cavity mode by a single particle, initially prepared in the trap ground state, as function of detuning between pump laser and cavity mode. Photon number (c) and its variance (d) as a function of the pump strength for different negative detunings. Here we have chosen the case of \nth{11} cavity mode as discussed above.}
\label{img:mode19-deltascans}
\end{figure}
To examine intrinsic resonances in the response of the coupled system to find the physically most interesting operating frequencies, we first probe the system by tuning the pump laser across the empty cavity resonance at low intensity. Compared to the empty cavity resonance we find a red shift for the effective resonance (fig.~\ref{img:mode19-deltascans}a), where its maximum as well as its position change nonlinearly with the pump strength (blue to green to red curve), indicating an increasingly efficient coupling by tighter particle localization at the anti nodes of the field. On the blue side of the effective resonance we get motional heating, leading to less effective particle-field coupling and line broadening. The variance of the photon number, fig.~\ref{img:mode19-deltascans}b, is greatly enhanced for stronger pumping, hinting at a strong deviation from a classical coherent intra-cavity field state (note the changed axis scale). The origin of these fluctuations will become much more obvious by looking at the field Q-function which we will study in more detail in the next section on the dynamics of the ordering transition. In fig.~\ref{img:mode19-deltascans}c we show the stationary photon number dependence of the intra-cavity field as a function of the pump amplitude for fixed detuning. Operating closely to cavity resonance on the red detuned side the transition from zero to finite field occurs at lower pump strength and much steeper (blue line) until we end up at the blue side of the effective resonance (orange line), where ordering is connected to heating and counteracts coherent scattering. Therefore, in the following we will always consider sufficient red detuning of the pump fields to ensure combined ordering and cooling.

\subsection{Single particle ordering dynamics}
As indicated by the frequency scans above and also found in classical models~\cite{griesser2010vlasov}, the instability of the trap ground state against light scattering is connected to a transition to an ordered phase only if we operate well on the cooling red side of the effective cavity resonance. Otherwise the particles are soon heated out of the ordered state, which was also observed experimentally in Z\"urich where a sufficiently large red detuning had to be used~\cite{baumann2010dicke}. For a numerical analysis of the transition from the weakly scattering trap ground state to an ordered radiating state, we will first start with a single particle and a single frequency model, which can be solved directly for a wide parameter range without too much numerical effort.

\begin{figure}[ht]
  \begin{subfigure}[b]{0.3\textwidth}
    \includegraphics[width=1\textwidth]{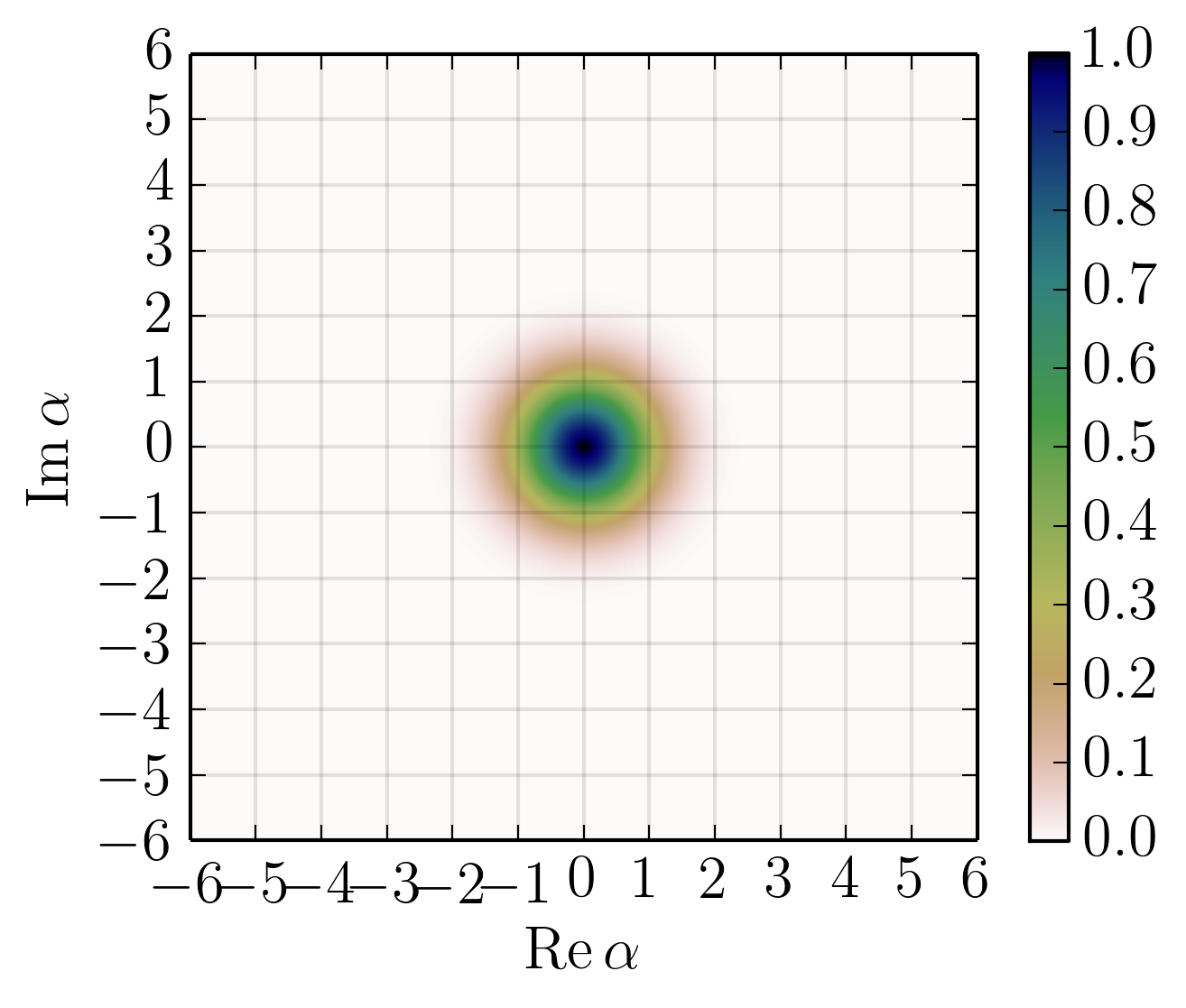}
    (a)
  \end{subfigure}
  \begin{subfigure}[b]{0.3\textwidth}
    \includegraphics[width=1\textwidth]{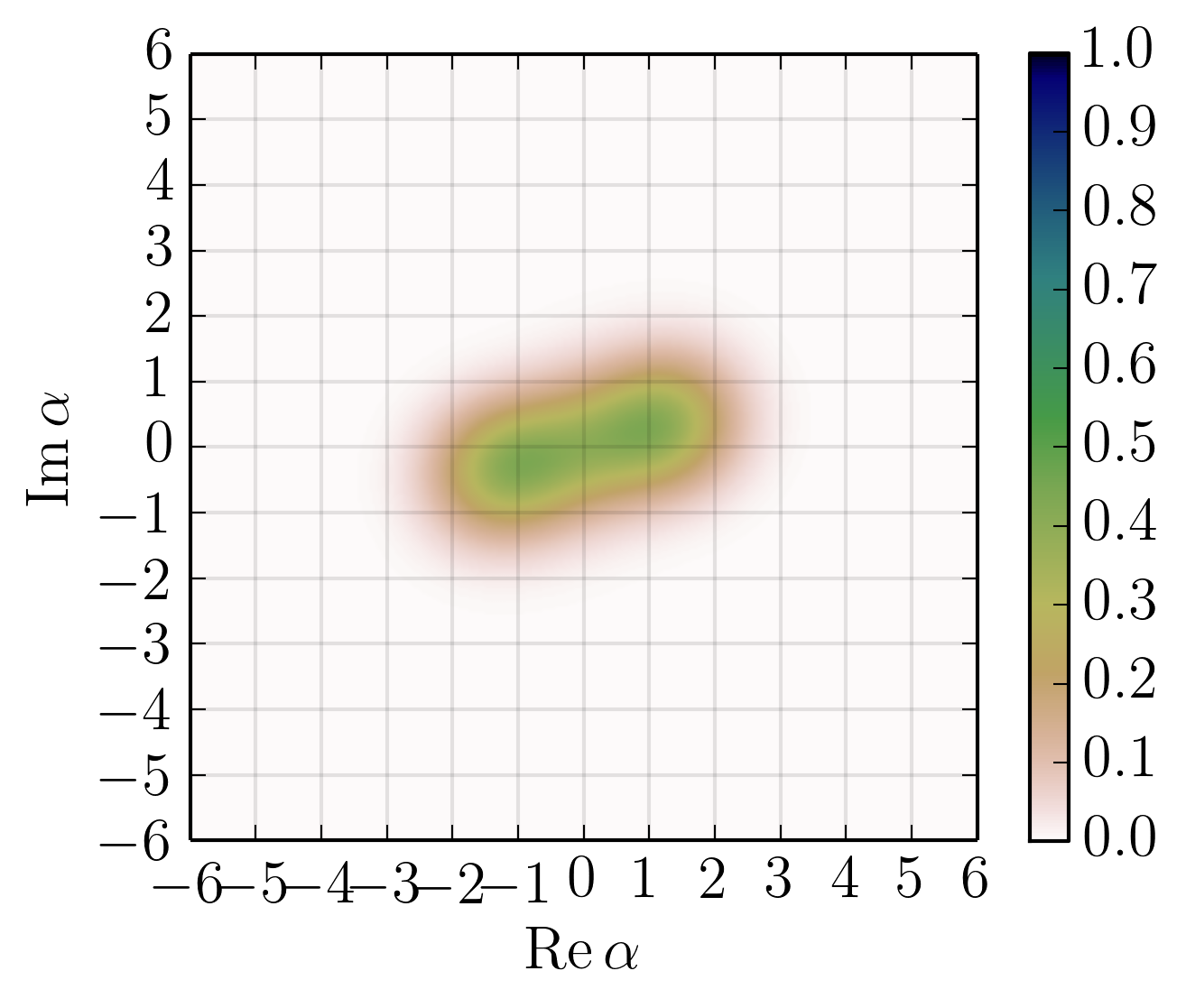}
    (b)
  \end{subfigure}
  \begin{subfigure}[b]{0.3\textwidth}
    \includegraphics[width=1\textwidth]{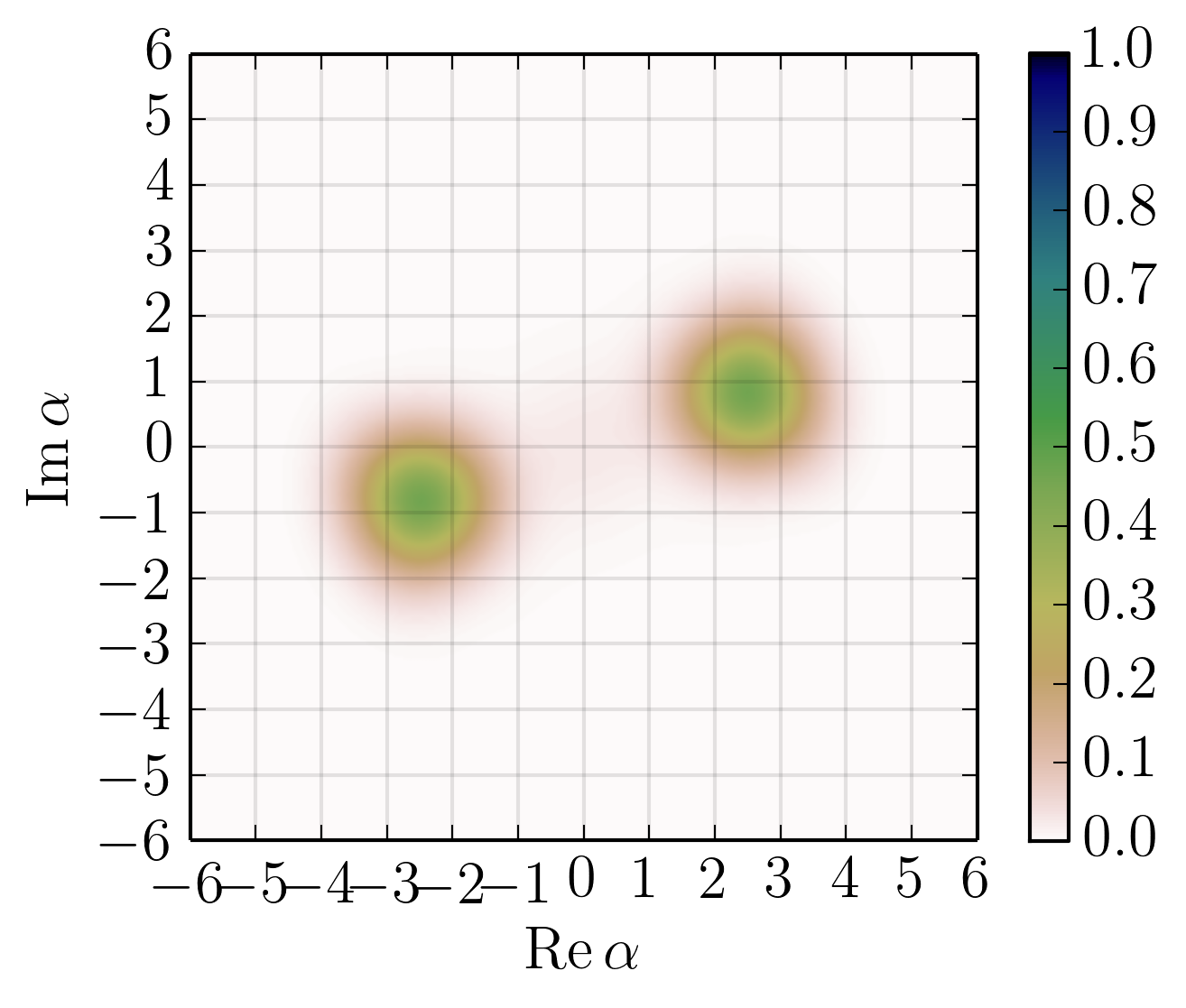}
    (c)
  \end{subfigure}
  \caption{Q-function of the steady state solution of the master equation where the \nth{19} cavity mode is pumped with $\eta=0\kappa$ (a), $\eta=1.5\kappa$ (b) and $\eta=4\kappa$ (c). Initially, the system is prepared with the particles in the trap ground state and with the cavity mode in the vacuum state. The detuning between pumping laser and cavity mode is set to $\Delta_c = -3 \kappa $ and $U_0 = -2 \kappa$.}
\label{img:mode19-qfuncs}
\end{figure}

\begin{figure}[ht]
  \begin{subfigure}[b]{0.3\textwidth}
    \includegraphics[width=1\textwidth]{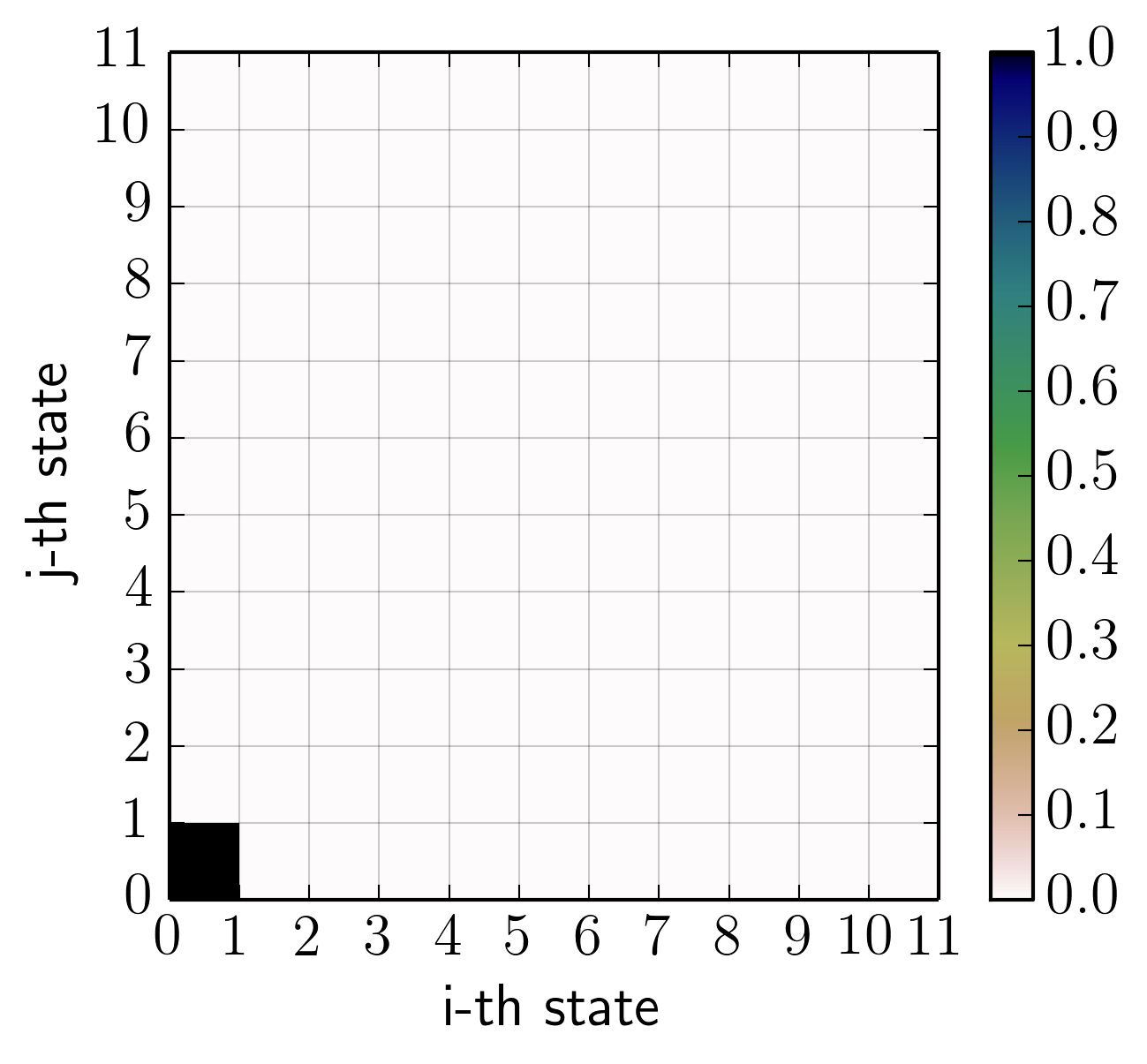}
    (a)
  \end{subfigure}
  \begin{subfigure}[b]{0.3\textwidth}
    \includegraphics[width=1\textwidth]{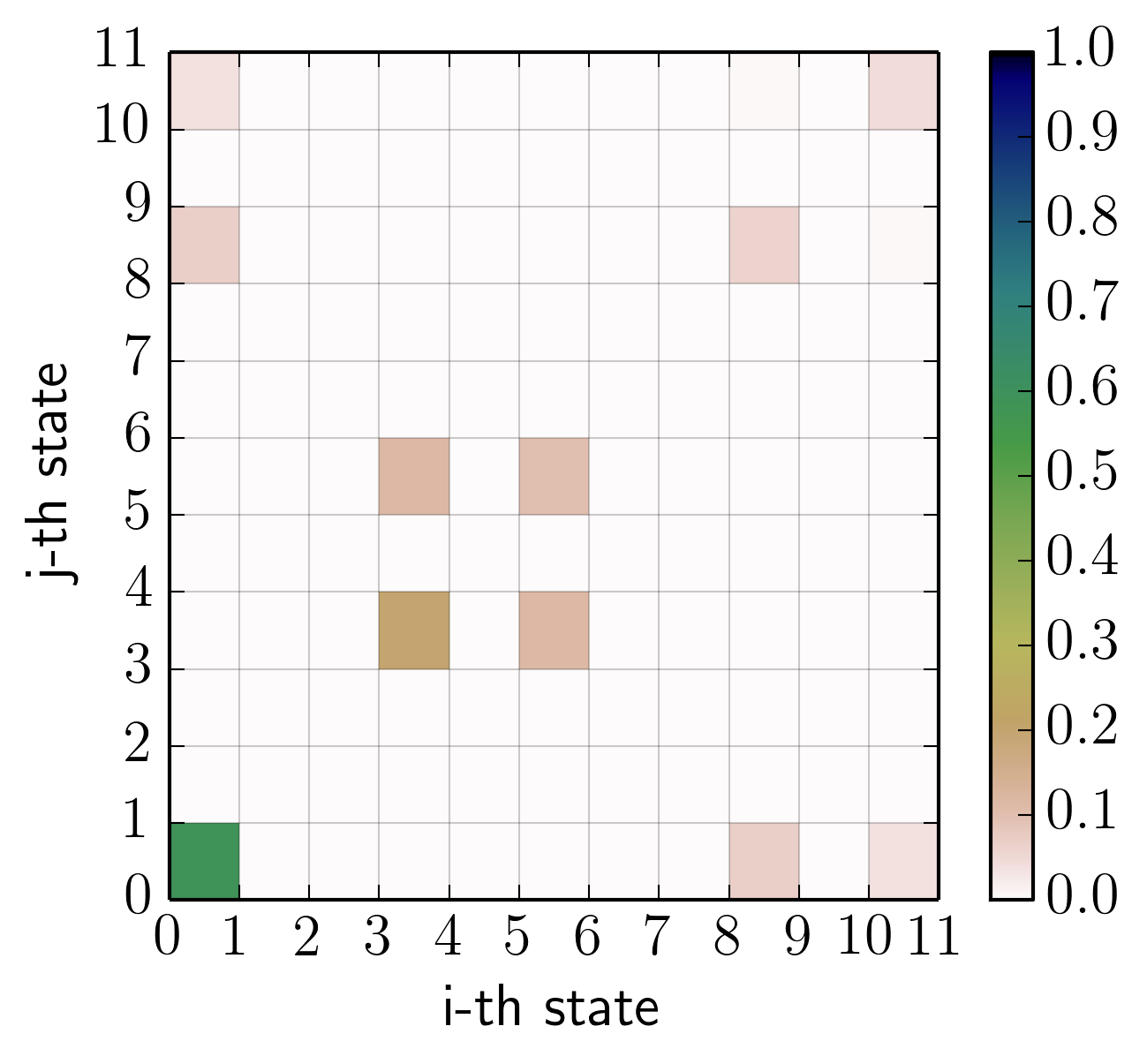}
    (b)
  \end{subfigure}
  \begin{subfigure}[b]{0.3\textwidth}
    \includegraphics[width=1\textwidth]{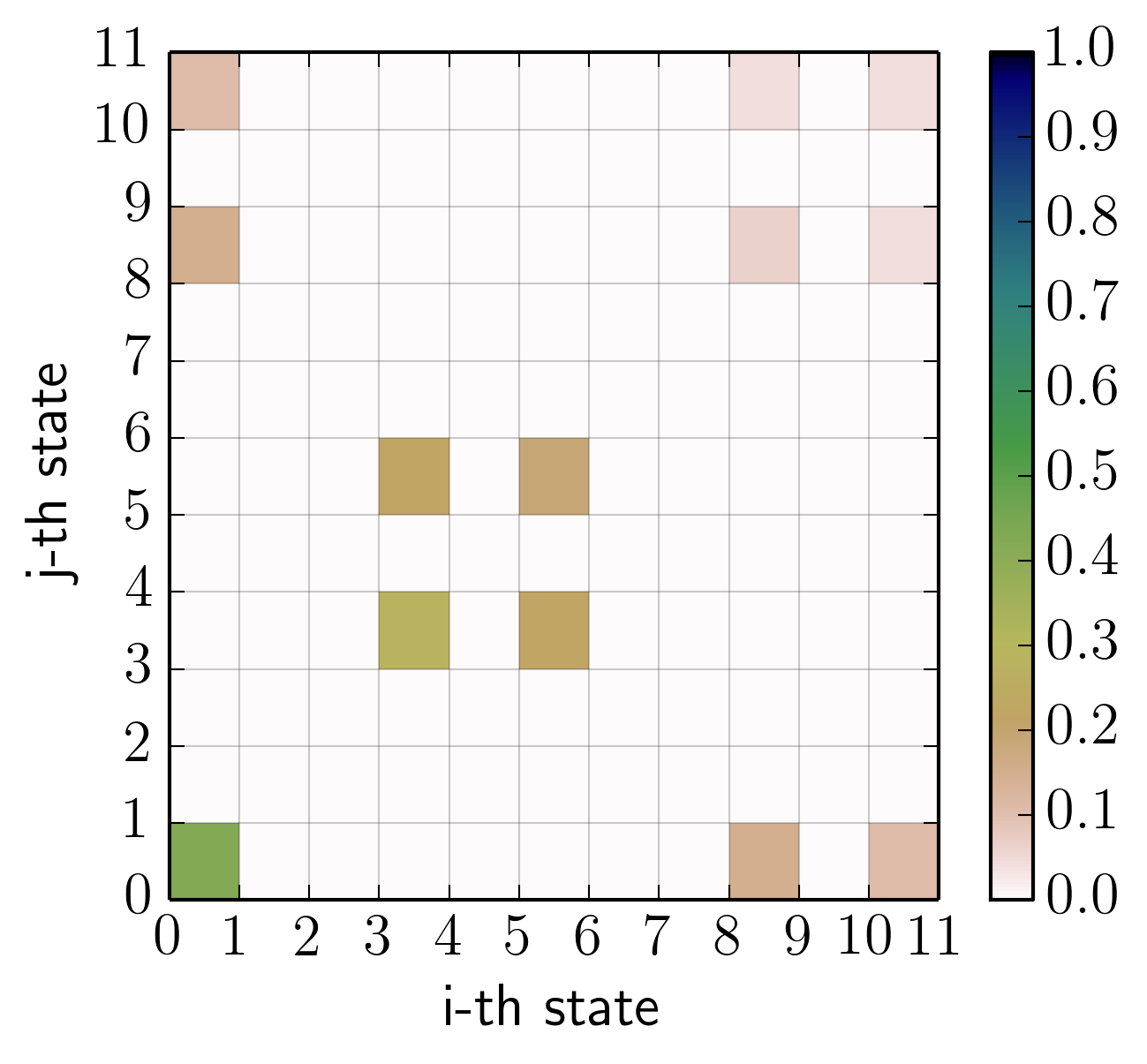}
    (c)
  \end{subfigure}
  \caption{Reduced particle density matrix in the trap eigenstate basis of the steady state solution of the master equation where the \nth{19} cavity mode is pumped with $\eta=0\kappa$ (a), $\eta=1.5\kappa$ (b) and $\eta=4\kappa$ (c). Initially, the system is prepared with the particles in the trap ground state and with cavity mode in the vacuum state. The detuning between pumping laser and cavity mode is set to $\Delta_c = -3 \kappa $ and $U_0 = -2 \kappa$.}
\label{img:mode19-DO}
\end{figure}

In the following we present some typical results. For suitably chosen detuning, i.e.\ $\Delta_c=-3\kappa$, we see a pronounced threshold for the pump amplitude, above which the number of scattered photons and also the corresponding fluctuations strongly increase (fig.~\ref{img:mode19-deltascans}). More detailed insight in the properties of the cavity field can be gained from the corresponding Q-functions shown in fig.~\ref{img:mode19-qfuncs} for three different pump strengths. The distribution changes significantly from a vacuum like distribution via a squeezed vacuum type shape to a clear bimodal phase space distribution, where the field is concentrated around two regions of opposite phase.
At the same time we find a concurrent fast grow of the population of the third and fifth excited trap states (fig.~\ref{img:mode19-DO}). For the strongest pump also higher trap modes, in this case the \nth{8} and \nth{10}, get substantially occupied. Thus, even for a moderate pumping power at a single pump frequency a Dicke-type two mode approximation will already fail.
\begin{figure}[ht]
  \begin{subfigure}[b]{0.3\textwidth}
    \includegraphics[width=1\textwidth]{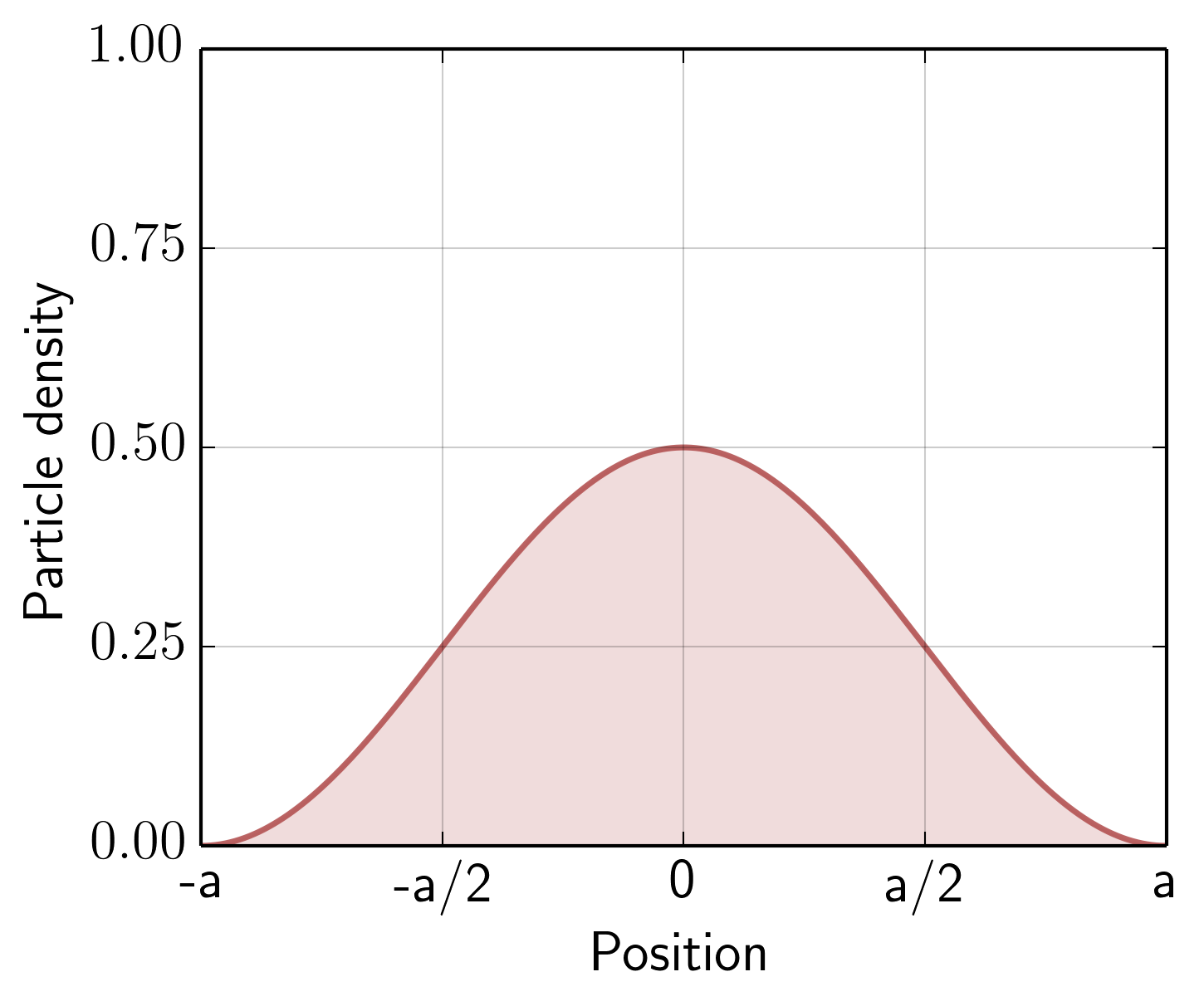}
    (a)
  \end{subfigure}
  \begin{subfigure}[b]{0.3\textwidth}
    \includegraphics[width=1\textwidth]{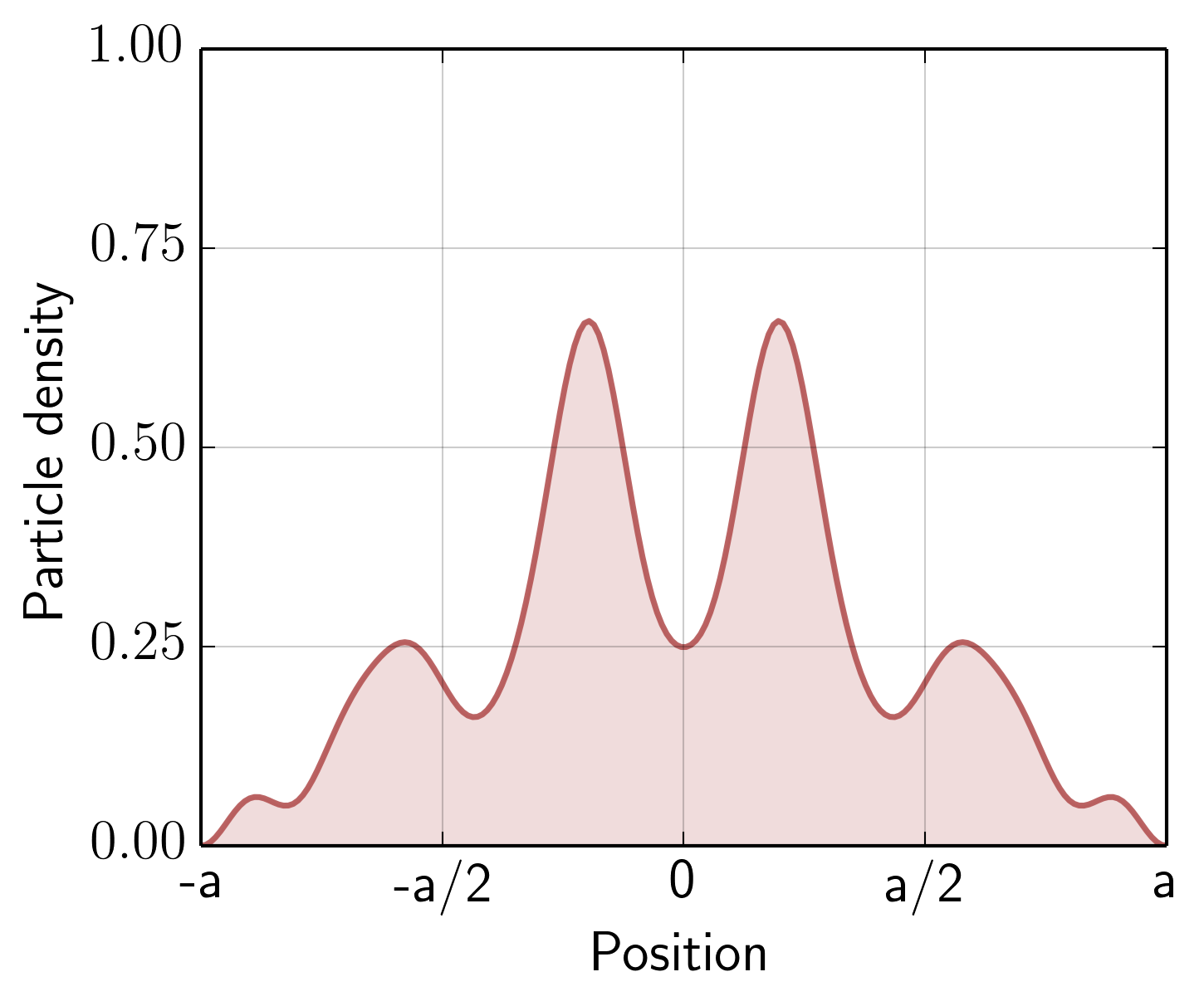}
    (b)
  \end{subfigure}
  \begin{subfigure}[b]{0.3\textwidth}
    \includegraphics[width=1\textwidth]{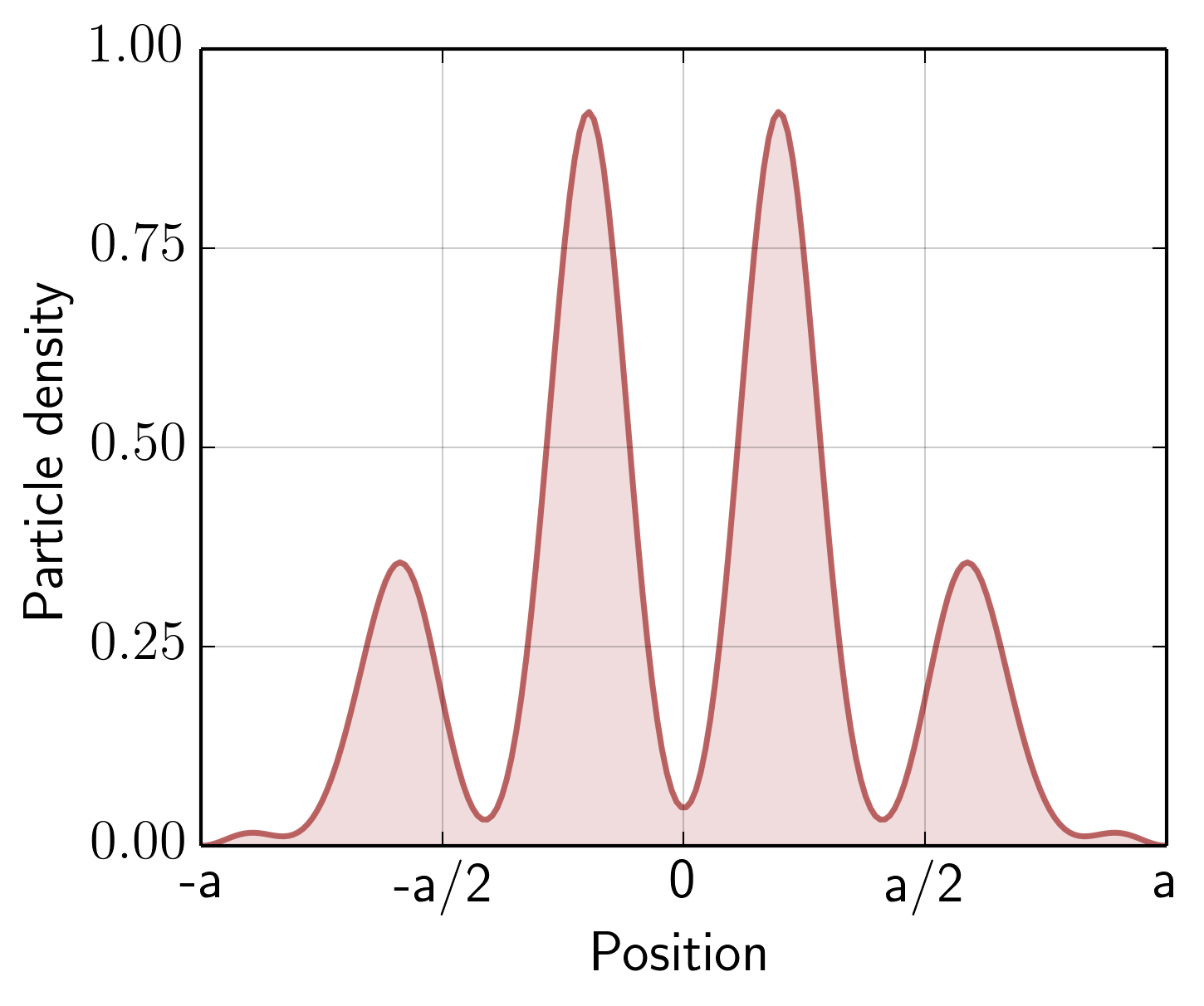}
    (c)
  \end{subfigure}
  \caption{Particle density of the steady state solution of the master equation where the \nth{19} cavity mode is pumped with $\eta=0\kappa$ (a), $\eta=1.5\kappa$ (b) and $\eta=4\kappa$ (c). Initially, the system is prepared with the particles in the trap ground state and with cavity mode in the vacuum state. The detuning between pumping laser and cavity mode is set to $\Delta_c = -3 \kappa $ and $U_0 = -2 \kappa$.}
\label{img:mode19-particledensity}
\end{figure}
Looking at the particle density, fig.~\ref{img:mode19-particledensity}, which exhibits a half wavelength modulation compared to the mode function depicted in fig.~\ref{img:mode19-conf}, one might at first expect no light scattering, as the overlap of the particle density with the positive and negative sections of the cavity mode amplitude are equal and thus the scattering should cancel. One possible explanation for the fact that there is indeed light scattered into the cavity, is that the master equation describes the evolution of stochastic averages and under this point of view we could interpret this density operator steady state as an average over trajectories where either only even antinodes or only odd antinodes are occupied, resulting in non-zero scattering for each trajectory. However, already in individual Monte Carlo trajectories every single node is occupied, therefore disproving this idea. Numerical analysis of the wave function of such an trajectory shows that it has two different parts, one that consists of a particle density concentrated on the odd antinodes, $\ket{x_+}$, connected with the cavity field $\ket{\alpha}$ and another part consisting of a particle density centered around the even antinodes, $\ket{x_-}$, connected to a cavity field $\ket{{-\alpha}}$. The MCWF trajectory in the limit $t\to\infty$ is well described by
\begin{equation}
  \ket{\psi_{t\to\infty}} = \frac{1}{\sqrt{2}} (\ket{x_+}\ket{\alpha} + \ket{x_-}\ket{{-\alpha}}),
\end{equation}
where the particle states as well as the value of $\alpha$ will fluctuate moderately among different trajectories and points in time. For higher pump strengths these fluctuations decrease significantly and the quality of this approximation increases substantially. The same numerical analysis performed for the steady state of the master equation reveals an analogous approximation of the form
\begin{equation}
  \rho_{t\to\infty} = \frac{1}{2} (\rho_{x_+} \otimes \ket{\alpha}\bra{\alpha} + \rho_{x_-} \otimes \ket{{-\alpha}}\bra{{-\alpha}}).
\end{equation}
Overall, we can reproduce several of the results obtained for the microscopic self-ordering of a single particle in two prescribed wells along the pump axis~\cite{vukics2007microscopic}.

\subsection{Two particle self-ordering}
\begin{figure}[ht]
  \begin{subfigure}[b]{0.24\textwidth}
    \includegraphics[width=1\textwidth]{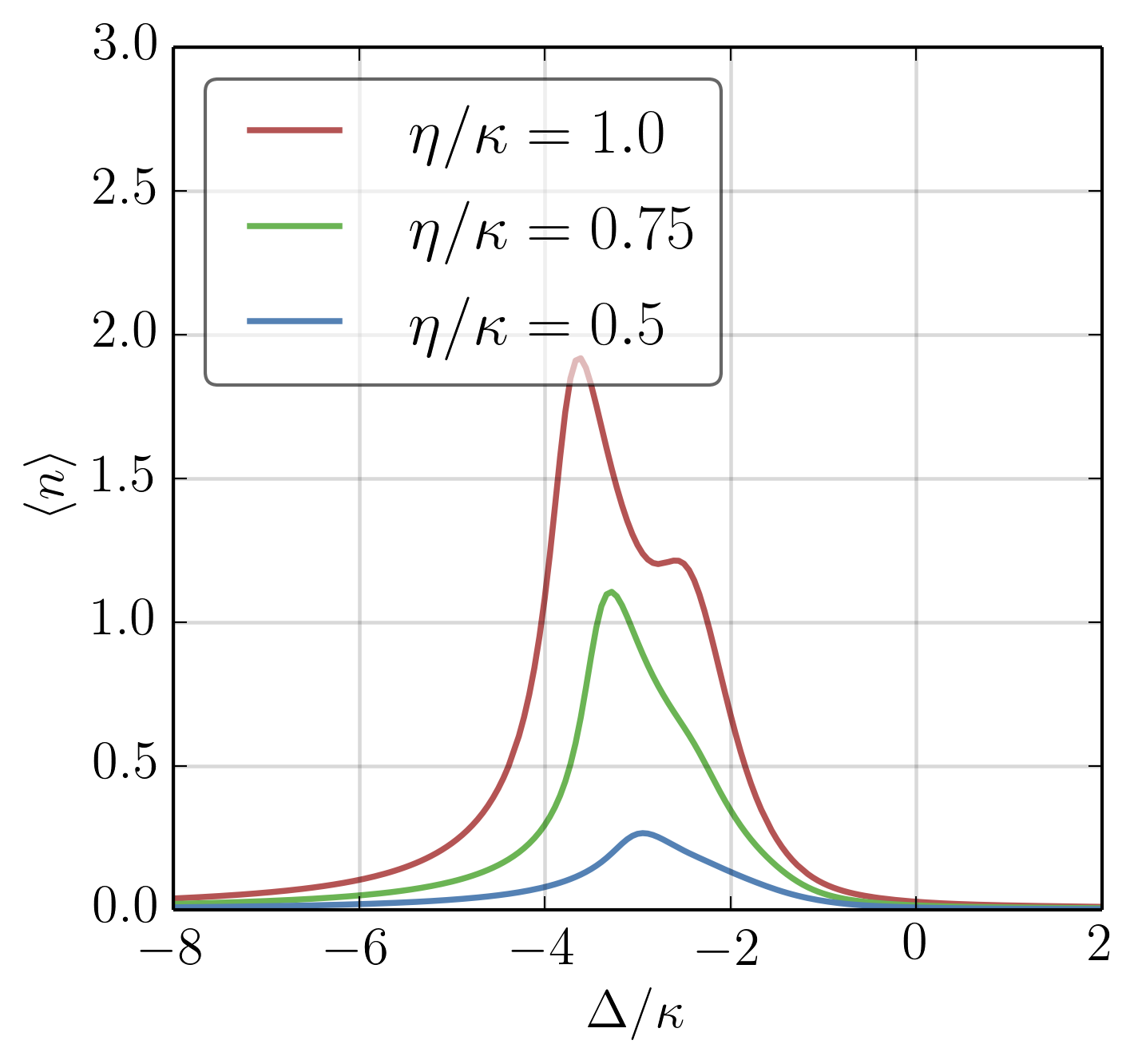}
    (a)
  \end{subfigure}
  \begin{subfigure}[b]{0.24\textwidth}
    \includegraphics[width=1\textwidth]{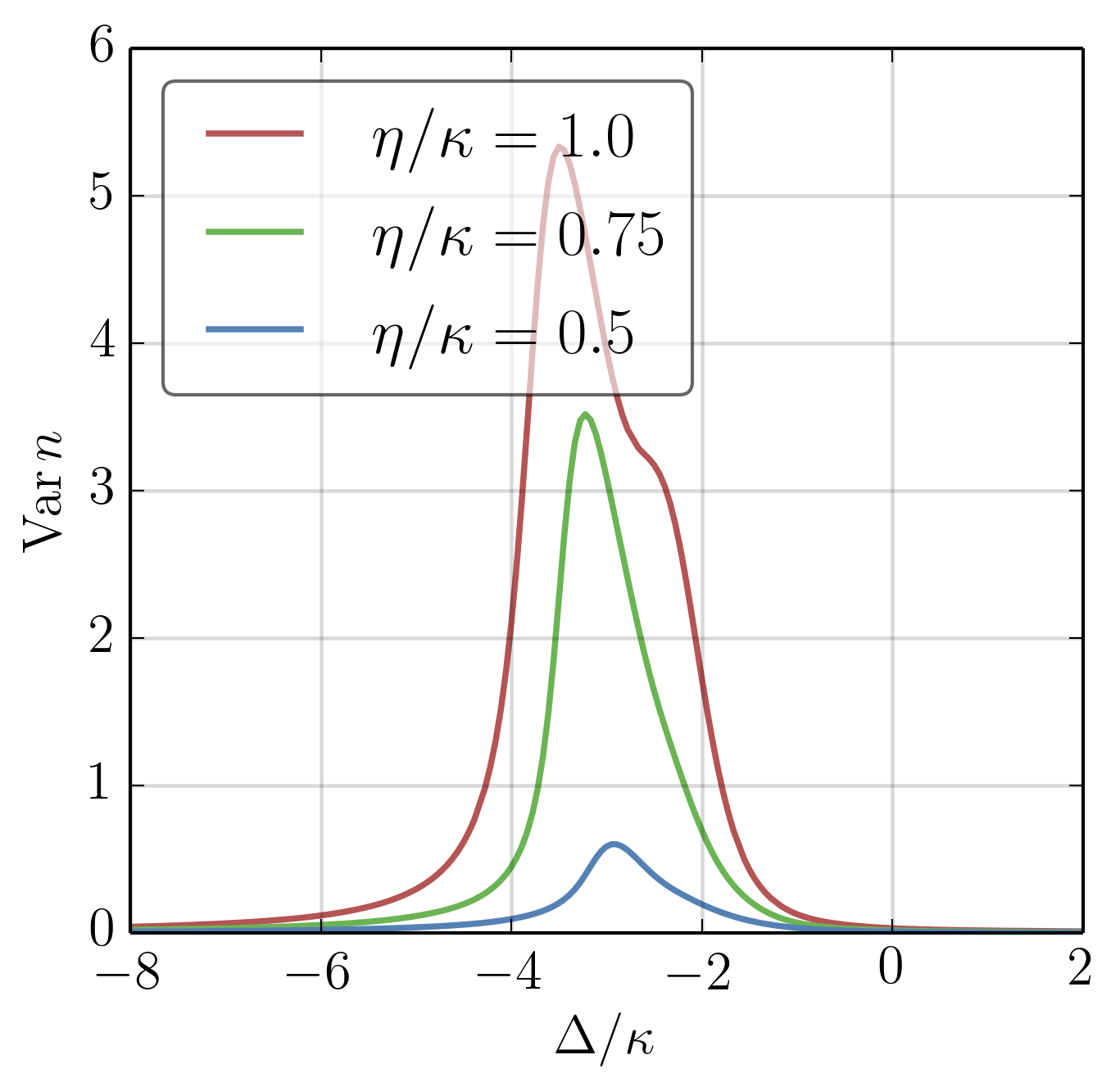}
    (b)
  \end{subfigure}
  \begin{subfigure}[b]{0.24\textwidth}
    \includegraphics[width=1\textwidth]{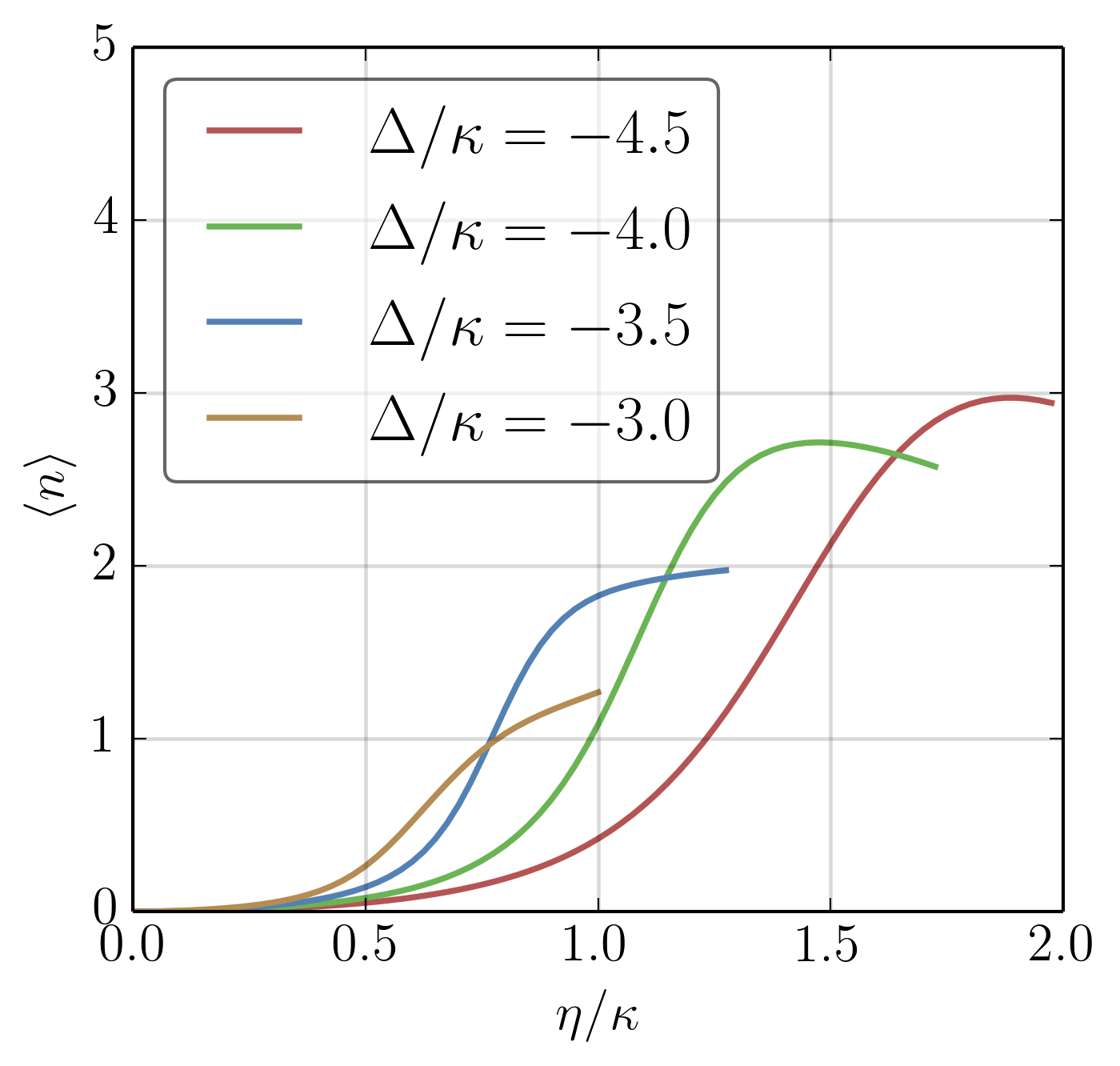}
    (c)
  \end{subfigure}
  \begin{subfigure}[b]{0.24\textwidth}
    \includegraphics[width=1\textwidth]{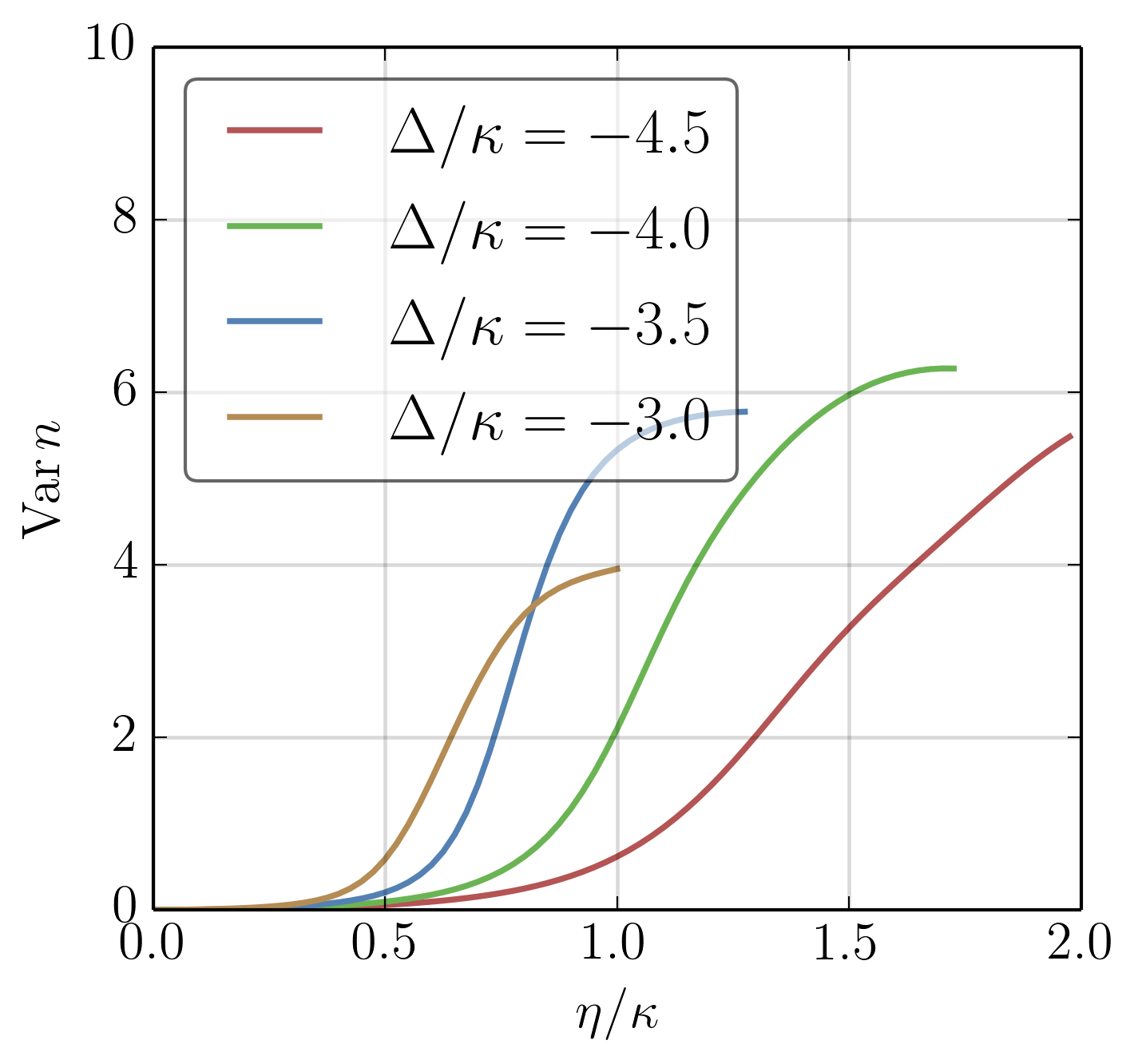}
    (d)
  \end{subfigure}
  \caption{Steady state expectation number of photons scattered into the cavity mode (a) and their variance (b) for two particles initially prepared in the trap ground state and different pump strengths as a function of the detunings chosen equal for both modes. In plots (c) and (d) we show the same quantities but as a function of the pump strength for different negative detunings.}
\label{img:mode11-2part-scans}
\end{figure}

\begin{figure}[ht]
  \begin{subfigure}[b]{0.3\textwidth}
    \includegraphics[width=1\textwidth]{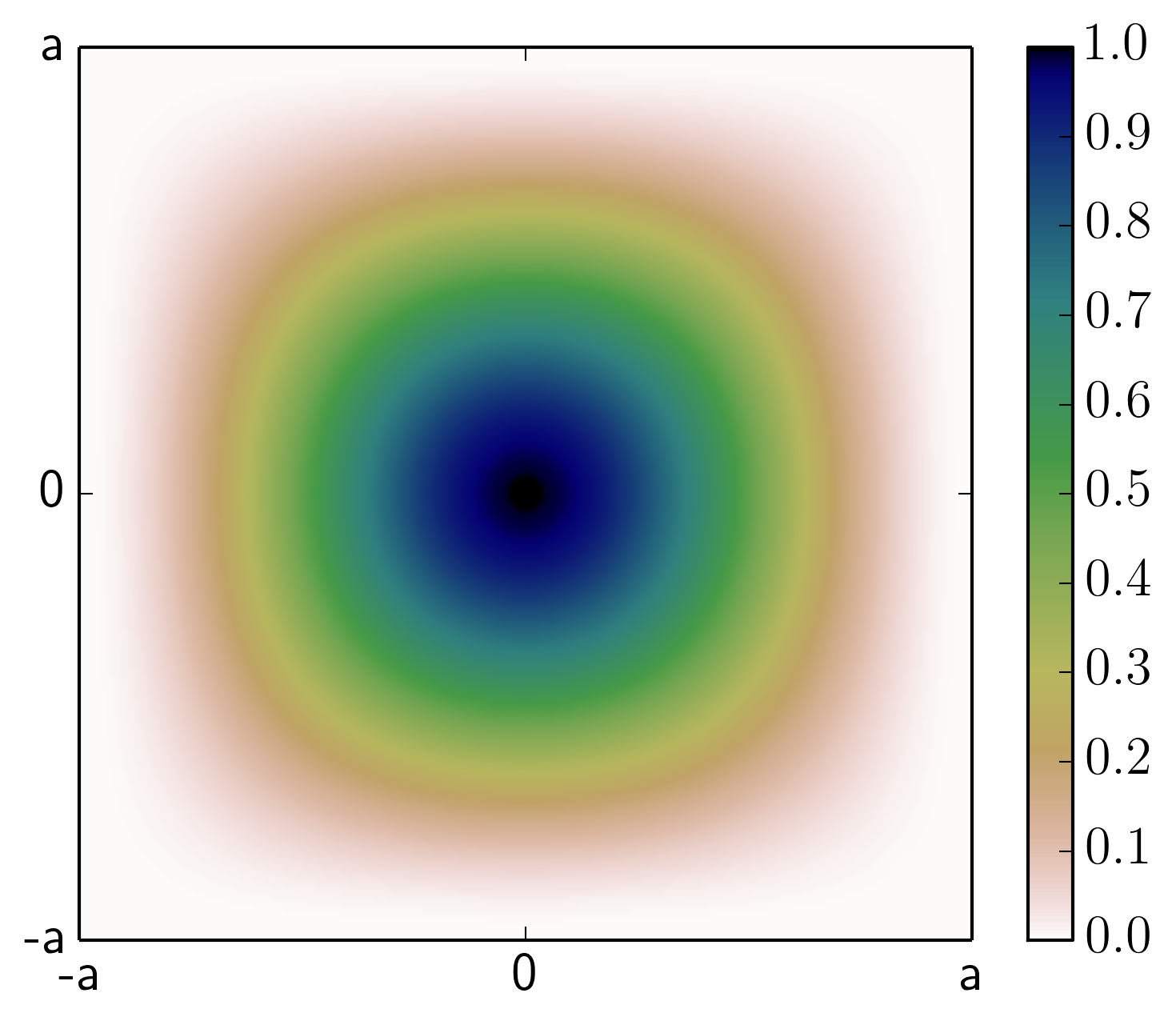}
    (a)
  \end{subfigure}
  \begin{subfigure}[b]{0.3\textwidth}
    \includegraphics[width=1\textwidth]{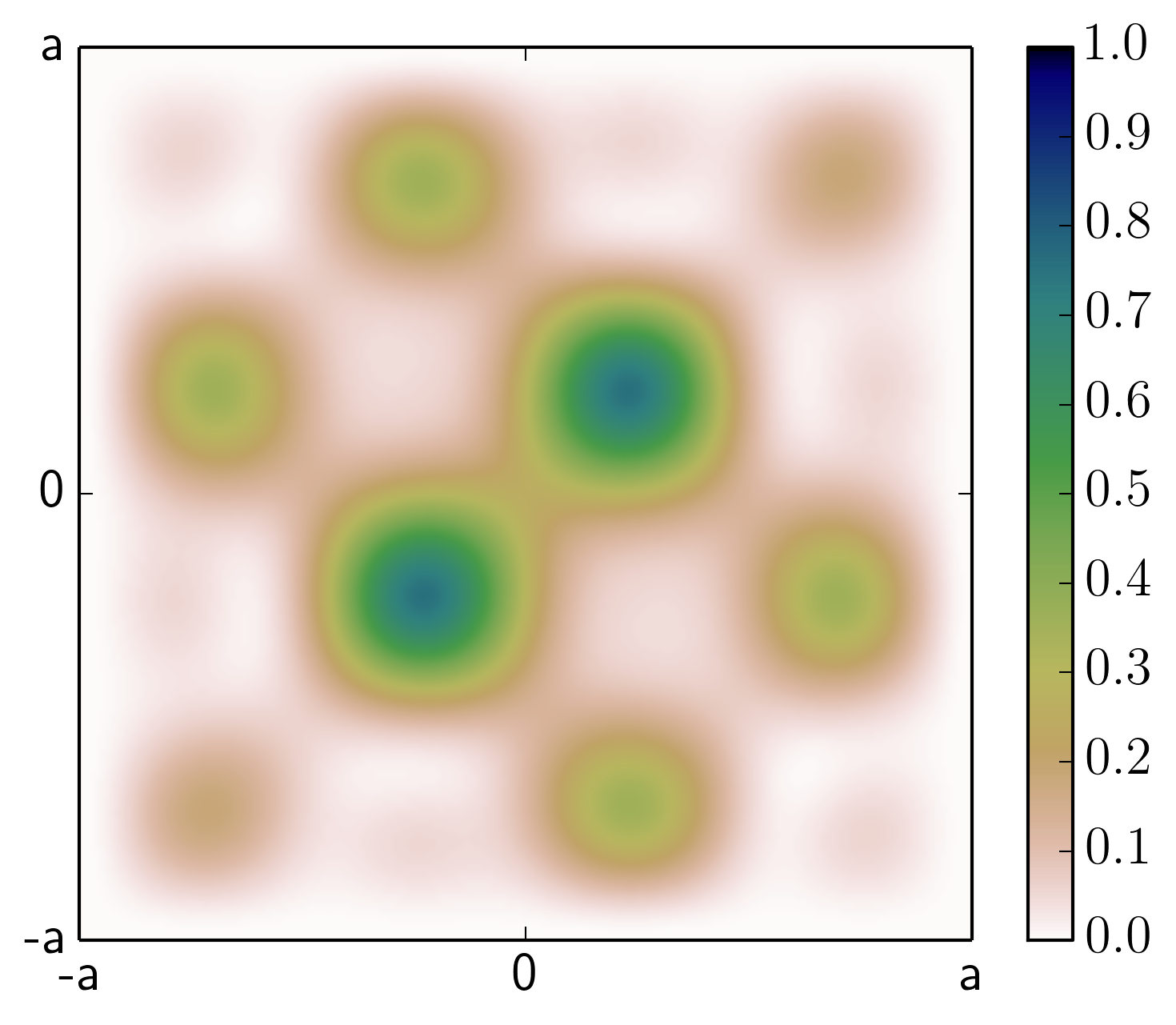}
    (b)
  \end{subfigure}
  \caption{Position density distribution $\rho_{x_1x_2}$ for 2 particles without pumping (a) and pumping of the \nth{19} cavity mode with $\eta=4\kappa$ and $\Delta_c = -3\kappa$ (b), revealing a clear preference of the particles to occupy the same wells, or at least wells with the same associated field state.}
\label{img:mode19-2particle-rhox1x2}
\end{figure}

With increasing particle numbers the accessible Hilbert space grows very fast and even without considering collisional interaction the time evolution becomes much more complex and numerically challenging compared to the single particle case. In general, the trapped particles will scatter light collectively and thus become correlated. Nevertheless, the average particle density, which is of central importance to light scattering, looks qualitatively similar and also the two-particle threshold and photon number fluctuations do not differ by much apart from a common shift and scaling, as indicated by fig.~\ref{img:mode11-2part-scans}. Extra peaks appearing for higher pump strengths in the resonance curve indicate different pair correlations of the atoms as discussed in some detail in~\cite{mekhov2007probing} and can be seen in the spatial density correlations visualized in fig.~\ref{img:mode19-2particle-rhox1x2}.
For adequately strong pump strengths, as depicted in fig.~\ref{img:mode19-2particle-rhox1x2}b, the diagonal elements are very pronounced, demonstrating a strong correlation between the positions of the particles. This implies that the system is not in a simple product state and thus goes again beyond a simple Dicke two mode model, which offers the same single motional mode for all particles.

\section{Simultaneous two color pumping}
As we have seen above, for a single, strong enough pump frequency the particles attain a rather complex superposition of trap eigenstates, which can be split approximately into two components corresponding to the two distinct field states of equal amplitude but opposite phase. In this section we now allow for two simultaneously injected excitation frequencies. Both frequencies are tuned closely to distinct longitudinal cavity modes, which are assumed to be separated sufficiently far, so that scattering of each field can occur solely into the corresponding mode. Each of the two laser powers and the detuning from their corresponding mode can be varied independently across the ordering threshold and for independent injection again has two possible stationary states. As one of our central goals in this work we now study key properties of the enriched dynamics for simultaneous operation of these pumps.

\subsection{Stationary state manifold for two color pump}
\begin{figure}[ht]
  \begin{subfigure}[b]{0.3\textwidth}
    \includegraphics[width=1.0\textwidth]{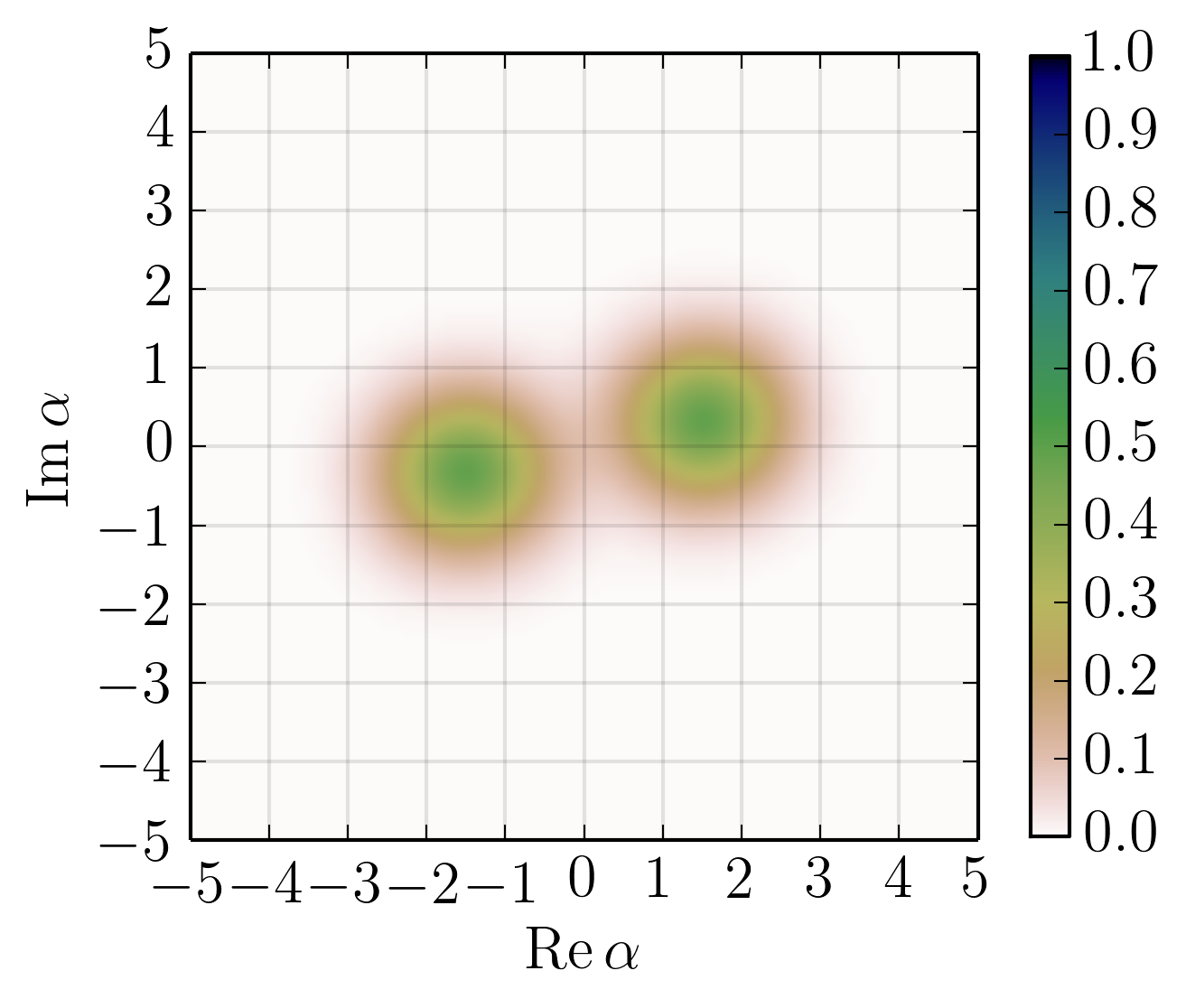}
    (a)
  \end{subfigure}
  \begin{subfigure}[b]{0.3\textwidth}
    \includegraphics[width=1.0\textwidth]{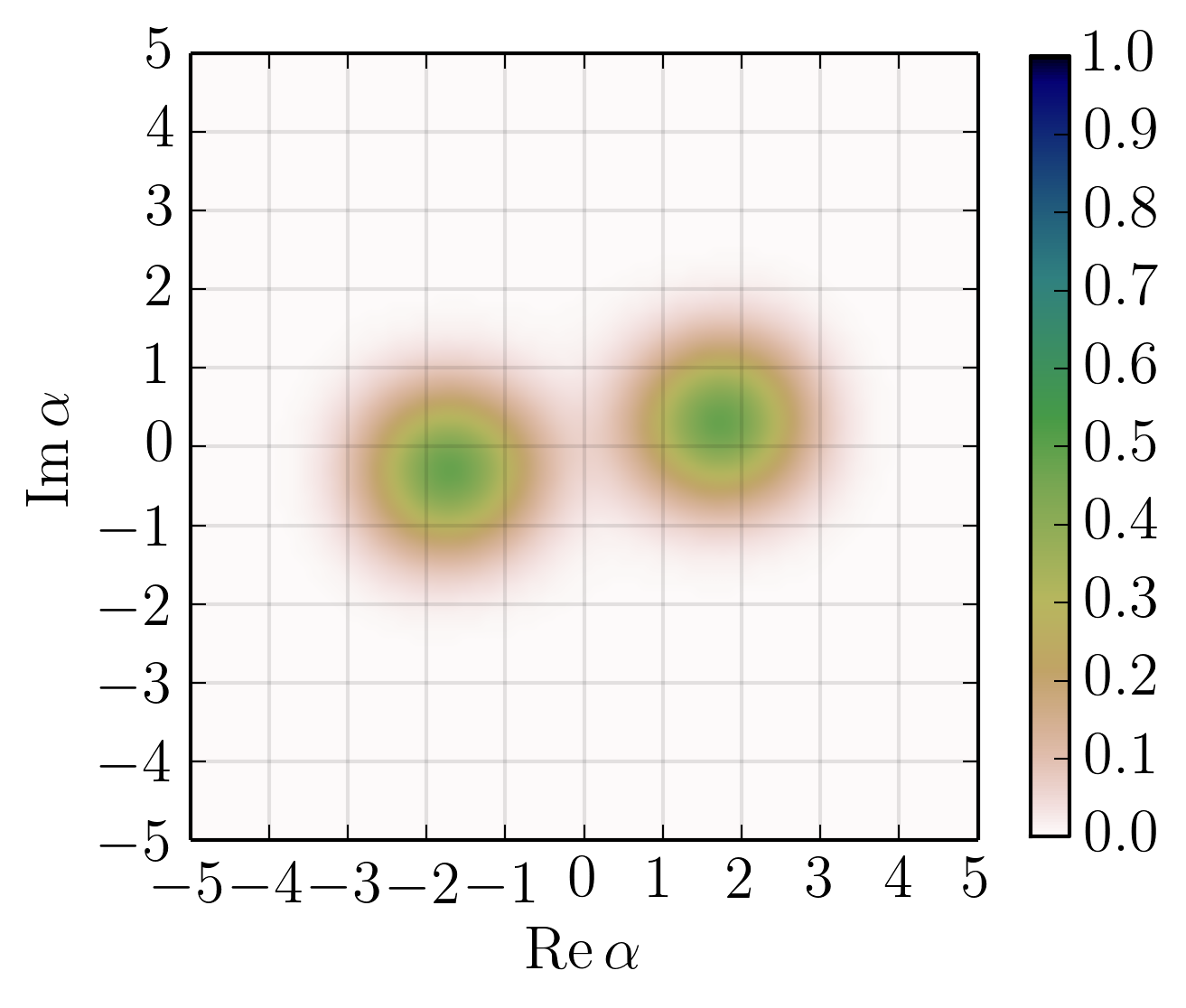}
    (b)
  \end{subfigure}
  \begin{subfigure}[b]{0.3\textwidth}
    \includegraphics[width=1.0\textwidth]{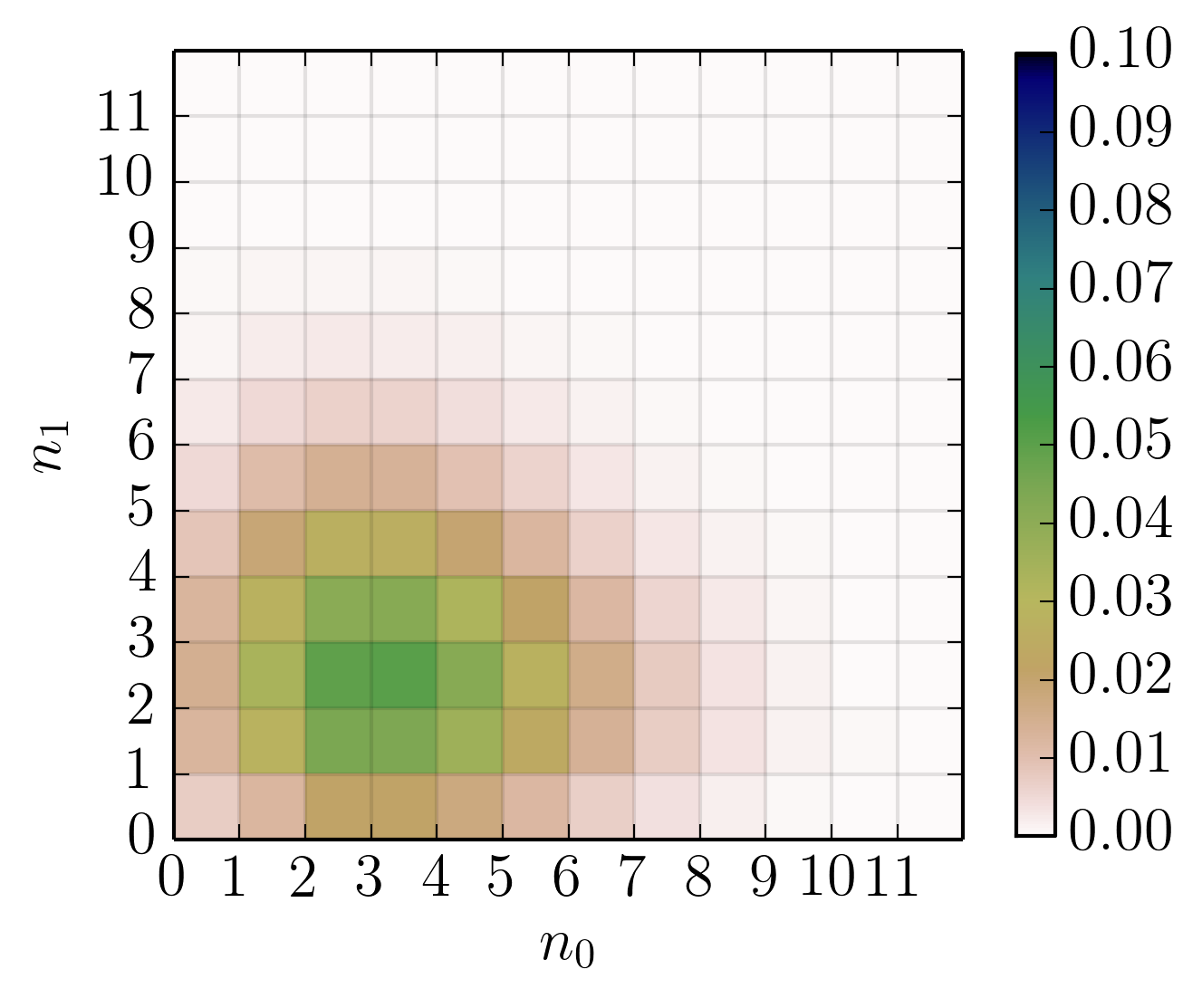}
    (c)
  \end{subfigure}
  \caption{(a) Averaged Q-function of mode 11, (b) Q-function of mode 19 and (c) joint photon number distribution probabilities $\rho_{n_1n_2}$ for two simultaneously strongly pumped modes where we have chosen $\eta_0=5\kappa,\eta_1=6\kappa$ and $\Delta_c^{11}=\Delta_c^{19}=-4\kappa$.}
\label{img:modes11-19-n0n1correlation}
\end{figure}

As a generic example we choose simultaneous operation on the \nth{11} and \nth{19} cavity mode and a trap size of 1/\nth{4} of the cavity length. This minimizes cross coupling matrix elements which facilitates an easier analysis of the complexity of the dynamics. As both lasers act on the same particles, we find a competition between the possible optimal orders for each field. It is now interesting to see for which pump ratios and strengths a single or even both orders are at least metastable and when the system is able to find completely new spatial distributions close to optimum for both fields. In general, besides a global optimum configuration several metastable local scattering maxima and potential energy minima exist. We will use QMCWF simulations, allowing us to study the multitude of quasi stationary solutions and also to retrieve important characteristics of the time evolution. As any momentary particular order leads to a characteristic scattering, we can use the photon number to quantify the corresponding order parameters.
\begin{figure}[ht]
  \begin{subfigure}[b]{0.24\textwidth}
    \includegraphics[width=1\textwidth]{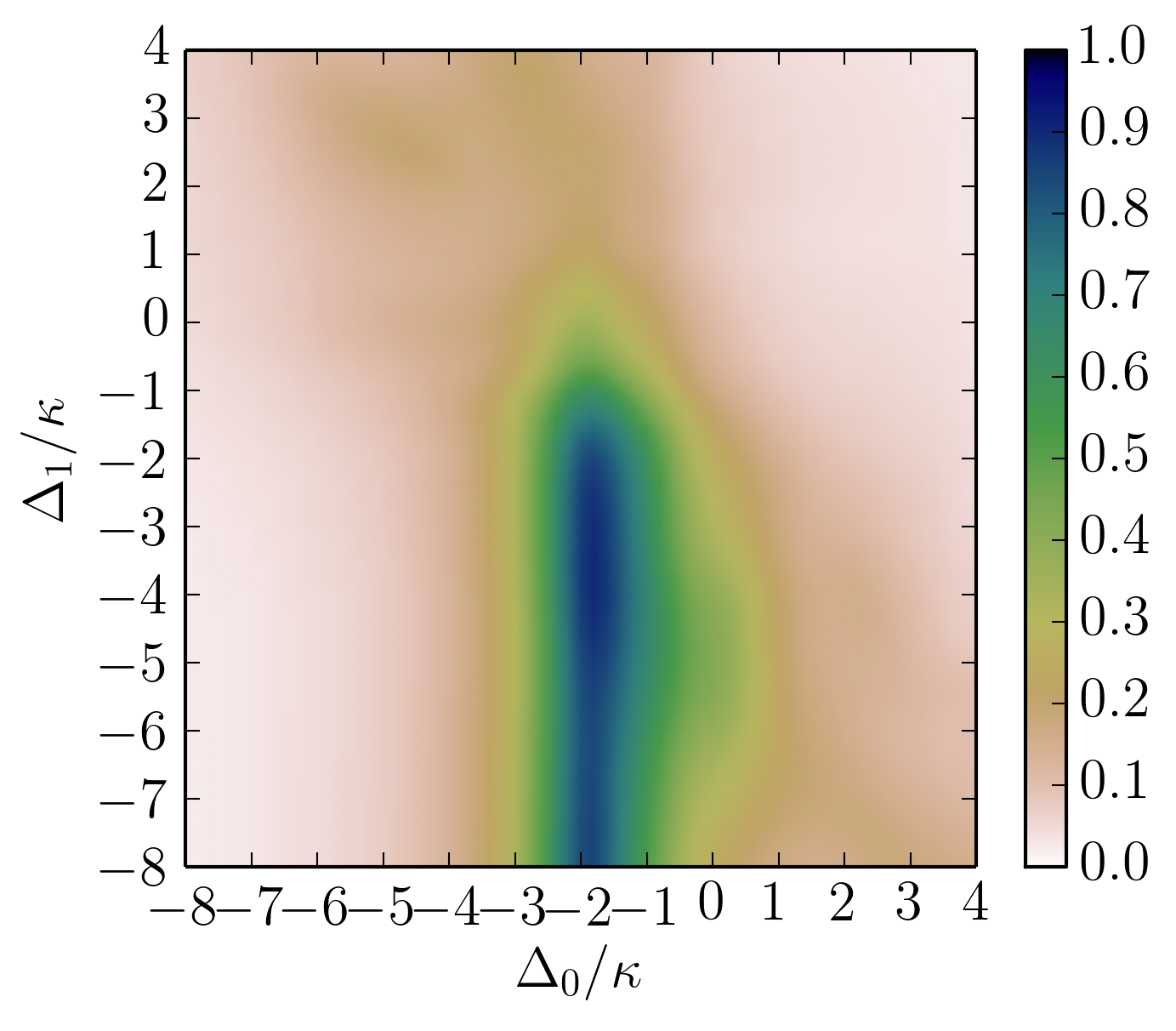}
    (a)
  \end{subfigure}
  \begin{subfigure}[b]{0.24\textwidth}
    \includegraphics[width=1\textwidth]{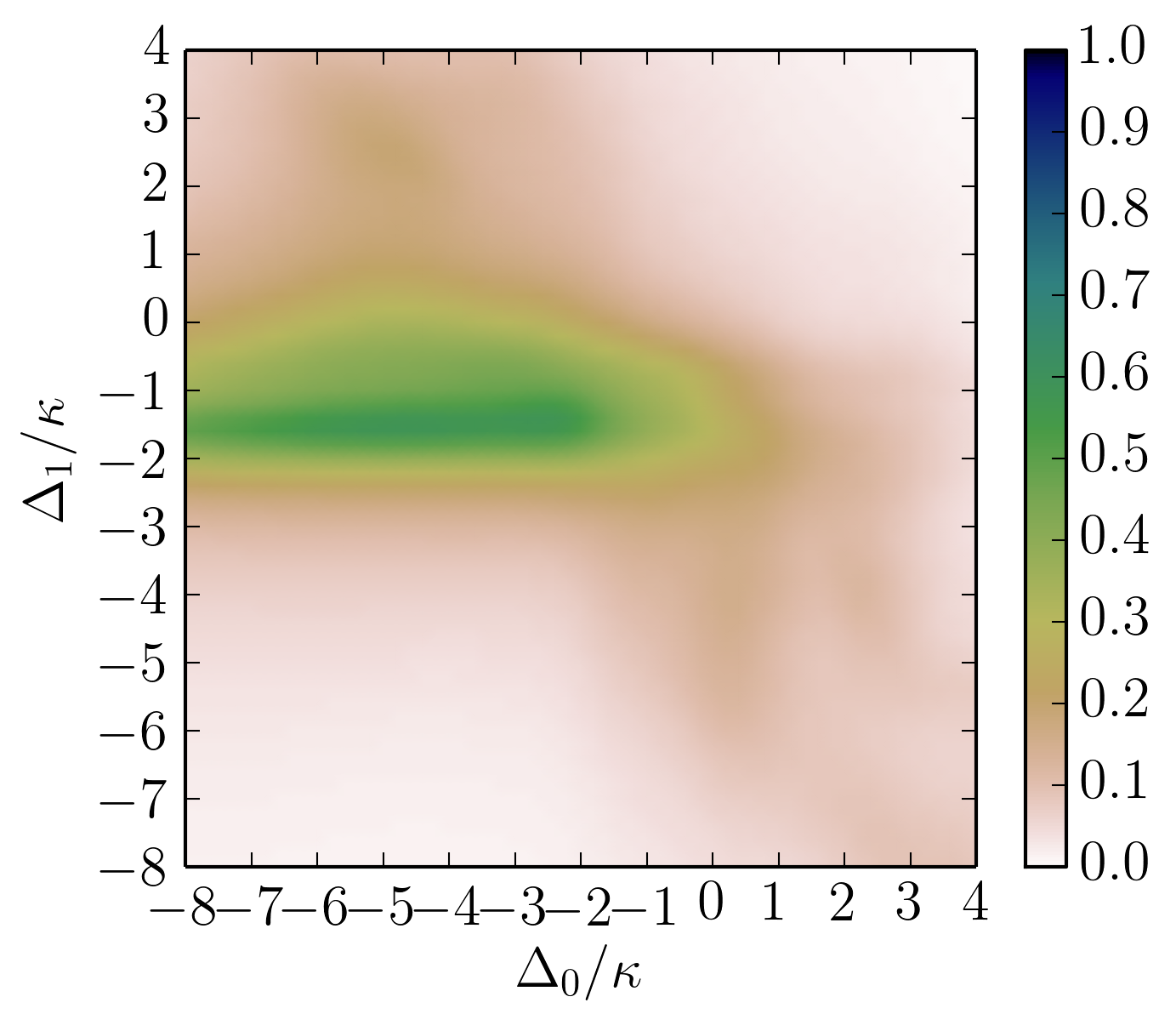}
    (b)
  \end{subfigure}
  \begin{subfigure}[b]{0.24\textwidth}
    \includegraphics[width=1\textwidth]{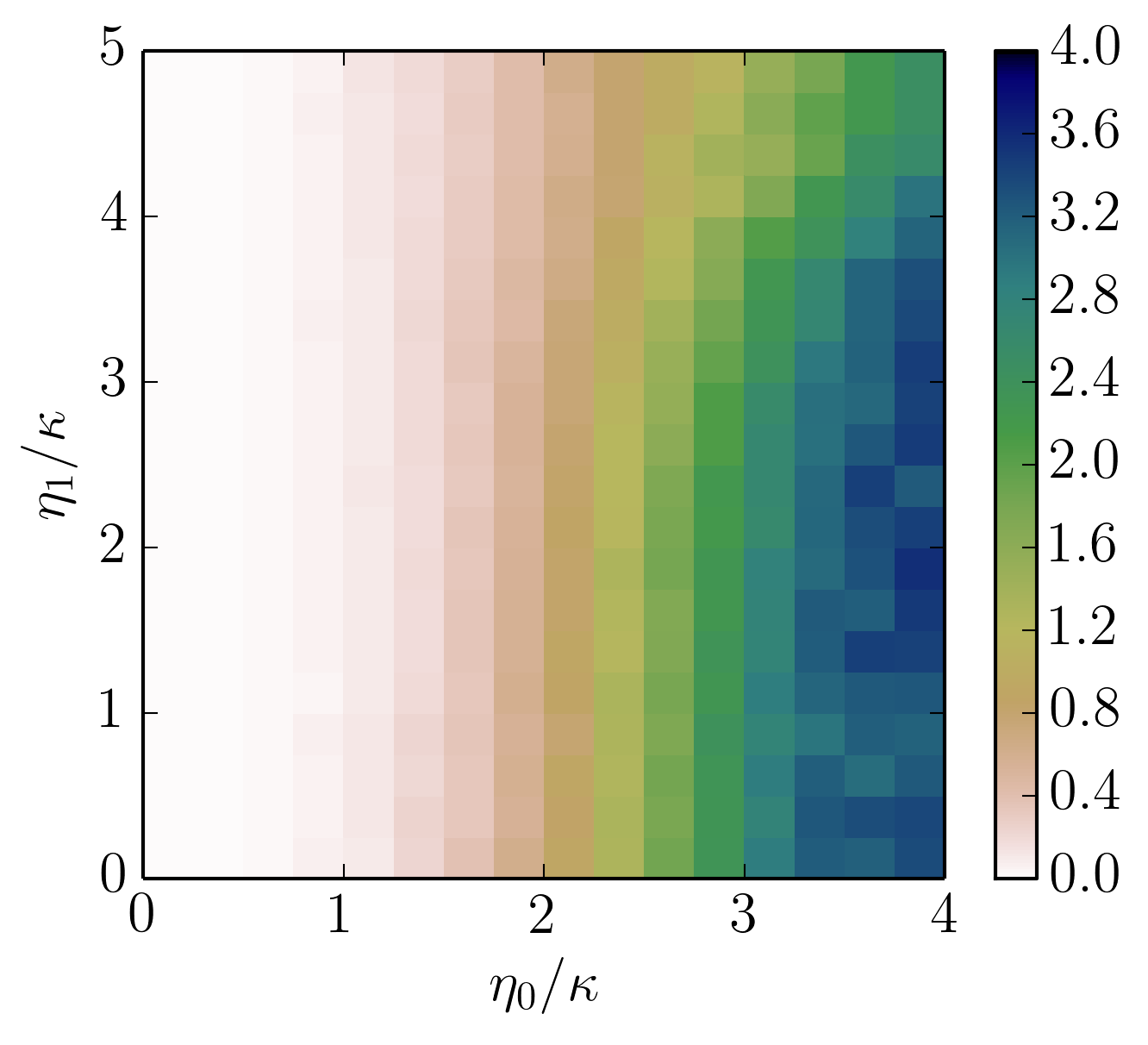}
    (c)
  \end{subfigure}
  \begin{subfigure}[b]{0.24\textwidth}
    \includegraphics[width=1\textwidth]{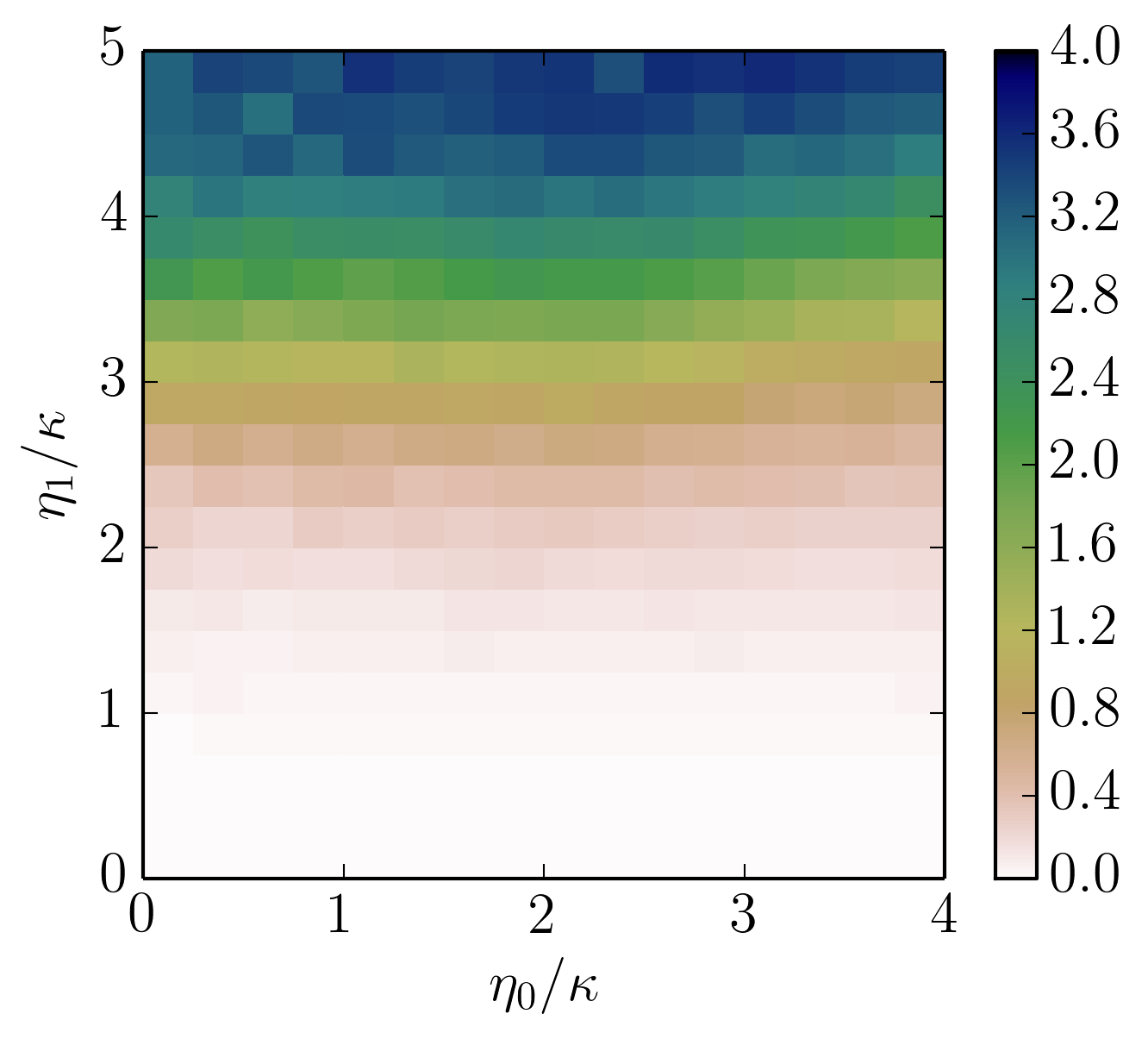}
    (d)
  \end{subfigure}
  \caption{Stationary photon number in the \nth{11} cavity mode (a) and \nth{19} cavity mode (b) as a function of the detunings $\Delta_c^{11},\Delta_c^{19}$ for equal pump amplitudes $\eta=1\kappa$. Stationary photon number in the \nth{11} cavity mode (c) and in the \nth{19} cavity mode (d) as a function of the pump amplitudes $\eta_{11}$ and $\eta_{19}$ for equal detunings $\Delta_c=-3\kappa$.}
\label{img:modes11_19-photonnumbers}
\end{figure}
To get a first insight, in fig.~\ref{img:modes11_19-photonnumbers} we plot the stationary averages of the photon numbers in the two modes, first as a function of the detunings for fixed pump amplitudes, then as a function of the two pump amplitudes for fixed negative detunings. While for preferential pumping of one field the corresponding photon number and a spatial order dominates, a finite pump at one field can either suppress or enhance the scattering and spatial order of the other mode. For our chosen example the threshold of the \nth{11} mode is shifted upwards by the field of the \nth{19} mode, while almost no effect is visible in the other direction. For close to equal pump strength, there are regions in parameter space where both modes oscillate simultaneously. This can be seen in the first two images of fig.~\ref{img:modes11-19-n0n1correlation}, showing the corresponding $Q$-functions. A bimodal distribution is also visible in the spatial density distribution shown in fig.~\ref{img:modes11-19-n0n1correlation}c. Note, that no direct light scattering between the modes is possible because their frequency difference is too large. As can be inferred from the missing vacuum contribution in the Q-functions and the dominance of the diagonal in the photon number distribution on the lower left, the two modes are oscillating at the same time. Microscopically this corresponds to an particle-field state of the approximate form
\begin{equation}
  \ket{\psi_{t\to\infty}} = \frac{1}{\sqrt{2}} \big(\ket{x_+}\ket{\alpha_0}\ket{\alpha_1} + \ket{x_-}\ket{{-\alpha_0}}\ket{{-\alpha_1}}\big).
\label{eq:2mode-approx-state}
\end{equation}
Each field amplitude pertains to a characteristic component of the particle density distribution, possibly allowing for various orthogonal combinations. Again, the number of such states grows rapidly with increasing particle number, so that even for such a small system size one quickly reaches the limits of classical computers. Experimentally, much higher particle and mode numbers seem easily possible, allowing to study and simulate Hamiltonians of the type of eq.~\ref{eq:model-hamiltonian-compact}.

\subsection{Time evolution between different metastable states}
So far we have concentrated on average, long time properties of the system to characterize the multitude of stationary quantum states and their intrinsic correlations. It is, however, almost equally interesting and important to study the time evolution and temporal fluctuations of the system. Due to limited space, we will restrict ourselves to a single typical example when the pump is tuned somewhat above the common oscillation threshold of the system. In this case several possible spatial configurations will compete and pure quantum fluctuations turn out to be sufficient to induce transitions between different spatial orders.
\begin{figure}[ht]
  \centering
  \includegraphics[width=1\textwidth]{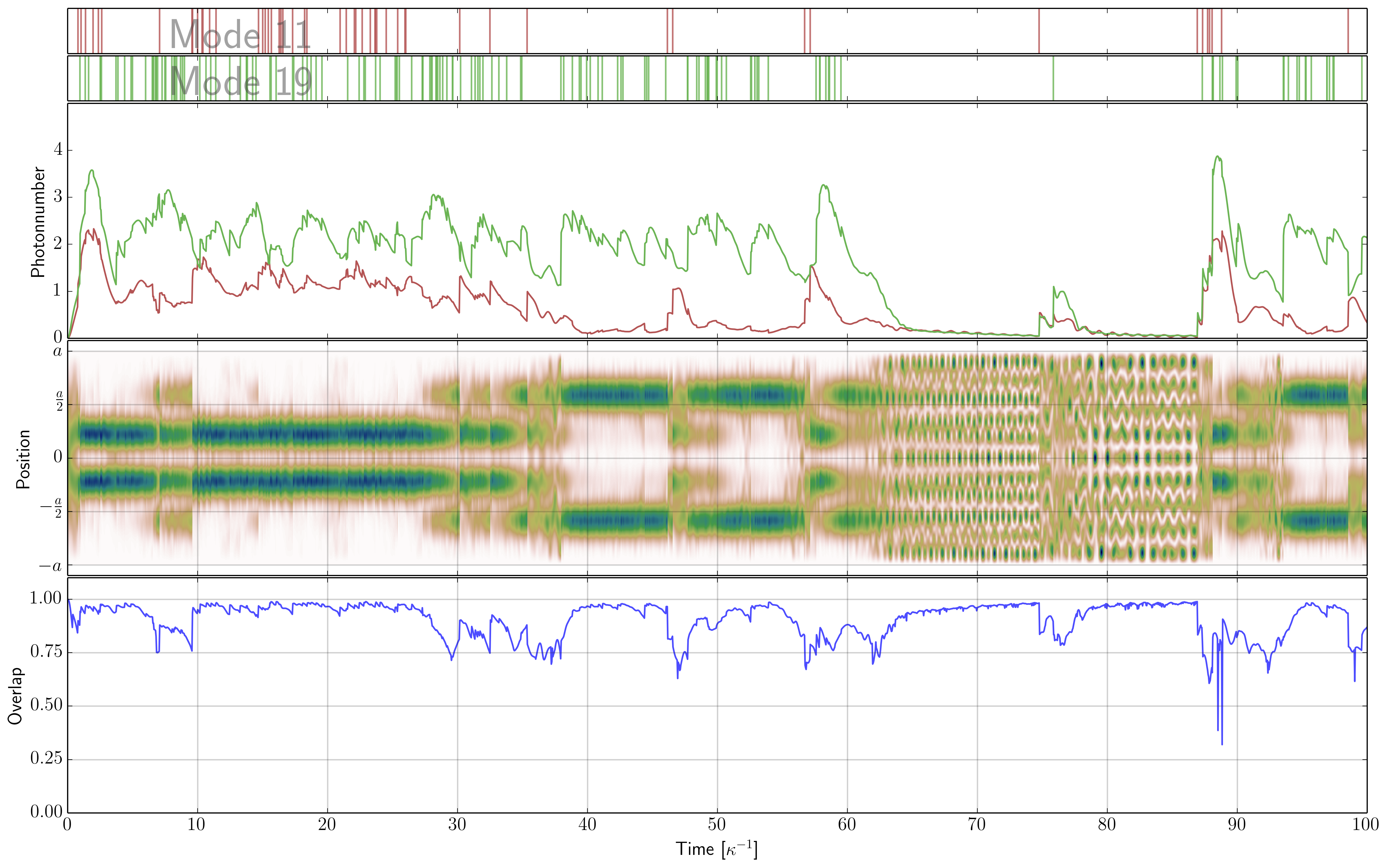}
  \caption{Single Monte Carlo wave function trajectory for a trapped particle with a two-color pump. The upper part shows intra cavity photon numbers and photo-detection events in the two modes, which correlates with the spatial density distribution of the particles below. In the last plot we show the momentary overlap with the heuristically expected approximate state of the form of eq.~\ref{eq:2mode-approx-state} with dynamically optimized state parameters.}
\label{img:modes11-19-timeevolution}
\end{figure}
This is presented in fig.~\ref{img:modes11-19-timeevolution} where we show the temporal behavior of a typical trajectory. Quantum jumps in the form of photo detection events eventually lead to a particle rearrangement between different spatial structures, in some cases even transitions back to a completely disordered states where light scattering stops almost completely can occur. In the ordered time periods we observe, as shown in the lowest graph, that the acquired state is very well approximated by a quantum state of the form of eq.~\ref{eq:2mode-approx-state}. The parameters of this state are dynamically obtained from the momentary maxima of the field Q-function. Note that each trajectory thus remains a coherent superposition between states of opposite field phase and the system does not break the inversion symmetry in respect to the trap center.
As state changes can be triggered by pure quantum fluctuations, one can of course actively switch between them by injecting corresponding signals into the cavity mode. Their stability can be increased by using higher pump powers. All these fascinating features are, unfortunately, computationally very time consuming and beyond the scope of this work at this point. Qualitatively, such effects could be investigated in strongly mode truncated models or by adiabatic elimination of the field dynamics.

\section{Conclusions}
Adding extra pump frequencies to the self-ordering dynamics of ultracold particles in a cavity field significantly changes the dynamics and enlarges the complexity of the system from a dynamical as well as computational point of view. Even with moderate particle and mode numbers, the limits of computability are soon reached. An alternative description in terms of an effective multi-mode Dicke model might catch essential qualitative features of the dynamics but quantitative reliable modeling is hardly possible. By including the trap dynamics in a consistent form we get a complex, widely tunable and very precisely controllable coupled quantum system, which can be mapped to a coupled oscillator system with tailorable long and short range interactions. The observation of photon scattering allows precise real time monitoring of the particle dynamics, which can then be controlled also by pump strength and detuning. With the help of frequency combs suitable cavity matched frequencies are available over a very wide range of longitudinal modes without the need for individual stabilization of each color. This should allow experimental studies far beyond Dicke physics as a very general quantum simulator.

In this work, we have mainly derived the basic Hamiltonian and only scratched the surface of the underlying complex nonlinear dynamics by means of a few simple but typical examples. A systematic investigation is certainly necessary and worthwhile but unfortunately beyond the scope of this work. Even more complex results could be expected from a generalization to fermionic particles~\cite{piazza2013umklapp} or particles with several internal levels~\cite{dalla2013keldysh}. Of course, one of the most interesting aspects and motivations of this approach is the fact that only minor additions to current experimental setups are required and the theoretical model for the current generations of experiments have proven to be surprisingly accurate and reliable~\cite{ritsch2013cold,gopalakrishnan2010viewpoint,baumann2010dicke,wolke2012cavity,sandner2013subrecoil}.

\section{Acknowledgments}
This work was supported by the Austrian Science Fund FWF project F4013 within the SFB FoQus. We also would like to thank Wolfgang Niedenzu for invaluable physical discussions.

\bibliographystyle{apsrev}
\bibliography{multimode_references}

\end{document}